
\input epsf

\newcount\figureno
\figureno = 1

%%%%%%%%%%%%%%%%%%%%%%%first set up fonts for captions%%%%%%%%%%%%%%%%%%%%%%%%

%use these fonts as john magnifies papers by magstep 1

\font\captionrmfont = cmr10     at 10 true pt
\font\captionitfont = cmti10    at 10 true pt
\font\captionslfont = cmsl10    at 10 true pt
\font\captionbffont = cmbx10    at 10 true pt
\font\captionttfont = cmtt10    at 10 true pt
\font\captiontitlefont = cmbx10 at 12 true pt

\font\captionmathfontrmb = cmr7   at 7 true pt
\font\captionmathfontrmc = cmr5   at 5 true pt
\font\captionmathfontita = cmmi10 at 10 true pt
\font\captionmathfontitb = cmmi7  at 7 true pt
\font\captionmathfontitc = cmmi5  at 5 true pt
\font\captionmathfontsya = cmsy10 at 10 true pt
\font\captionmathfontsyb = cmsy7  at 7 true pt
\font\captionmathfontsyc = cmsy5  at 5 true pt
\font\captionmathfontexa = cmex10 at 10 true pt
\font\captionmathfontbfb = cmr7   at 7 true pt
\font\captionmathfontbfc = cmr5   at 5 true pt

\def\setupcaptionfonts{
  \def\rm{\fam=0 \captionrmfont}
  \def\it{\fam=\itfam \captionitfont}
  \def\sl{\fam=\slfam \captionslfont}
  \def\bf{\fam=\bffam \captionbffont}
  \def\tt{\fam=\ttfam \captionttfont}

  \textfont0 = \captionrmfont
        \scriptfont0 = \captionmathfontrmb
        \scriptscriptfont0 = \captionmathfontrmc 
  \textfont1 = \captionmathfontita
        \scriptfont1 = \captionmathfontitb
        \scriptscriptfont1 = \captionmathfontitc 
  \textfont2 = \captionmathfontsya
        \scriptfont2 = \captionmathfontsyb
        \scriptscriptfont2 = \captionmathfontsyc
  \textfont3 = \captionmathfontexa
        \scriptfont3 = \textfont3
        \scriptscriptfont3 = \textfont3

  \textfont\itfam = \captionitfont
  \textfont\slfam = \captionslfont
  \textfont\bffam = \captionbffont
        \scriptfont\bffam = \captionmathfontbfb
        \scriptscriptfont\bffam = \captionmathfontbfc
  \textfont\ttfam = \captionttfont

  \rm
}

%%%%%%%%%%%%%%%%%%%%%%%%%%%%%%%%caption macro%%%%%%%%%%%%%%%%%%%%%%%%%%%%%%%%

\long\def\insertcaption#1{{
   \setupcaptionfonts
   \setbox0=\vbox{\hsize 0.8\hsize
      \noindent\captiontitlefont Fig. \number\figureno\
      \captionrmfont #1
      }

   \centerline{\box0}

   \global\advance\figureno by 1
   }}

%%%%%%%%%%%%%%%%%%%%%%%%%%%%%%%figure macros%%%%%%%%%%%%%%%%%%%%%%%%%%%%%%%%

\long\def\insertweirdfig[#1,#2]#3 {{

  \bigskip
  \vbox{
    \epsfysize = #2  
    \centerline{\hbox{\epsffile{#1}}}
    \medskip
    \insertcaption {#3}
    }
  \bigskip
  
}}

\long\def\insertfig[#1]#2 {{

  \bigskip
  \vbox{
    \epsfxsize = 0.9\hsize 
    \centerline{\hbox{\epsffile{#1}}}
    \medskip
    \insertcaption {#2}
    }
  \bigskip
  
}}

\long\def\inserttwofigs[#1,#2]#3#4 {{

  \epsfxsize = 0.45\hsize

%  \bigskip
  \hbox{
    \vbox{\hsize 0.5\hsize
      \epsfxsize = 0.9\hsize
      \centerline{\hbox{\epsffile{#1}}}
      \medskip
      \insertcaption {#3}
      }
    \vbox{\hsize 0.5\hsize
      \epsfxsize = 0.9\hsize
      \centerline{\hbox{\epsffile{#2}}}
      \medskip
      \insertcaption {#4}
      }
    }
%  \bigskip
  
}}

% Local Variables: 
% mode: plain-tex
% TeX-master: t
% End: 

\input phyzzx
\FRONTPAGE
\line{\hfill BROWN-HET-1021}
\line{\hfill August 1996}
\vskip0.5truein
\titlestyle{{THE STRUCTURE OF COSMIC STRING WAKES}}
\medskip
\author{A. Sornborger$^{1)}$, R. Brandenberger$^{2)}$, B. 
Fryxell$^{3)}$, and K. Olson$^{3)}$}
{$^{1)}$ DAMTP, University of Cambridge, Cambridge CB3 9EW, 
UK} \nextline
\indent {$^{2)}$ Physics Department, Brown University, Providence, RI  
02912, USA} \nextline
\indent {$^{3)}$ Insitute for Computational Science and Informatics,
George Mason University, Fairfax VA 22030 \foot{Postal Address: NASA
GSFC, Code 934, Greenbelt, MD 20771}}
\medskip
\abstract 

The clustering of baryons and cold dark matter induced by a single
moving string is analyzed numerically making use of a new
three-dimensional Eulerian cosmological hydro code$^{1)}$ which is
based on the PPM method to track the baryons and the PIC method to
evolve the dark matter particles.

A long straight string moving with a speed comparable to $c$ induces a
planar overdensity (a``wake").  Since the initial perturbation is a
velocity kick towards the plane behind the string and there is no
initial Newtonian gravitational line source, the baryons are trapped
in the center of the wake, leading to an enhanced baryon to dark
matter ratio. The cold coherent flow leads to very low post--shock
temperatures of the baryonic fluid.

In contrast, long strings with a lot of small-scale
structure (which can be described by adding a Newtonian gravitational
line source) move slowly and form filamentary objects.  The large
central pressure due to the gravitational potential causes the baryons
to be expelled from the central regions and leads to a relative
deficit in the baryon to dark matter ratio. In this case, the velocity of
the baryons is larger, leading to high post-shock temperatures.

Note: mpeg simulations may be found at
http://www.damtp.cam.ac.uk/user/ats25/

\chapter{Introduction}

The cosmic string theory has emerged as a promising model to explain
the origin of structure in the Universe$^{2)}$.  The primordial power
spectrum predicted by the model is scale-invariant, in reasonable
agreement with the results from recent observations of large-scale
structure.  Other predictions of the model which can be made based on
the linear theory of cosmological perturbations (e.g. the amplitude of
temperature anisotropies in the cosmic microwave background on COBE
scales) agree with observations to within the observational and
theoretical error bars$^{3)}$.

However, nonlinear effects in the cosmic string model are more
important and important at earlier times than is the case e.g. in
inflation-based models.  This is due to the fact that the seed
perturbations are nonlinear {\it ab initio} and that their
distribution is non-Gaussian.  This renders the study of the theory
much more complicated than the study of inflation-based models where
the initial perturbations are given by a Gaussian random field with a
small initial amplitude.

There are many uncertainties in our present understanding of the
cosmic string theory of structure formation.  To begin with, while it
is known that the distribution of strings obeys a scaling
solution$^{4)}$ which says that the statistical properties of the
string network are constant in time when all lengths are scaled to the
Hubble radius, the specific properties of this scaling solution are
not known.  A further uncertainty concerns the so-called small-scale
structure of the strings.  Are strings which cross a Hubble volume
straight or wiggly on smaller scales?  The answer to this question
influences both the mean velocity of the string and the gravitational
effects on the surrounding matter.  Finally, very little is known
about the nonlinear evolution of the string-induced mass
perturbations.  Since -- at least in a model in which the dark matter
is hot -- most of the matter goes nonlinear in large-scale
structures$^{5)}$ (of the scale of the Hubble radius at $t_{eq}$, the
time of equal matter and radiation), it is not possible to compute the
predicted galaxy and cluster properties and quantify their nonrandom
distribution without an understanding of the nonlinear dynamics.

There are various means of making progress towards a better
understanding of the predictions of the string model.  One approach is
to start from the output of high resolution simulations of the cosmic
string dynamics and to use the resulting string networks to provide
the initial conditions for large-scale $N$-body
simulations$^{6)}$. One difficulty with this approach is that it is
not easy to correctly take into account compensation, i.e. the fact
that at the time of the string producing phase transition, the energy
density fluctuations due to strings are exactly compensated by
fluctuations in the radiation energy density. Maybe more importantly,
the present cosmic string simulations$^{7)}$ do not have the
resolution to include all of the small-scale structure which is
generated by the nontrivial evolution of the strings.  It may even be
that the Nambu action on which the evolution of the strings is based
neglects some crucial physics$^{8)}$.  Finally, it is difficult to
investigate the dependence of the results of the $N$-body simulations
on the uncertainties in the input string distribution.

In this work, we follow an alternative approach.  We focus on the
evolution of structure induced by a single string.  To begin with,
this gives us better resolution for studying the nonlinear dynamics on
scales where nonlinear effects are crucial.  More importantly,
however, our approach will allow us to systematically investigate the
dependence of the results of the nonlinear evolution on the
uncertainties of the input physics.  We can study the dependence on
variables such as the small-scale structure of strings, the string
velocity, and perturbations in the matter surrounding the strings, to
name just a few.

The objective of our study is to analyze the clustering of dark matter
and baryons induced by cosmic strings.  Since according to recent
cosmic string evolution simulations$^{7)}$ most of the mass is in
strings which are long (relative to the Hubble radius), we here focus
on such strings.  An extension of our study to cosmic string loops is
not hard.  Since we are ultimately interested in comparing the
predictions of the string model to observations from optical and
infrared large-scale redshift surveys of galaxies, it is important to
track the evolution of dark matter and baryons separately.  Hence,
cosmological hydro-simulations are required.  The questions we
initially address in this paper are the determination of dark matter
and baryon profiles of cosmic string wakes and the dependence of the
results on the amount of small-scale structure on the strings.  In
future work, we intend to analyze and compare mechanisms which induce
the fragmentation of wakes into substructures.

Understanding the nonlinear dynamics of dark matter and baryons
induced by a single string will allow us at a later stage to combine
these results with an analytical toy model of the string scaling
solution to obtain information about the correlation  of galaxies and
galaxy clusters in the string model.

The outline of this paper is as follows: In Section 2 we give a brief
review of the features of the cosmic string model which are relevant
for our study.  Next, we discuss the methods used in this work: the
hydrodynamical equations, the basic numerical techniques and their
implementation.  In Section 4 we describe our simulations and present
the results.  We study planar collapse induced by a rapidly moving
string without small-scale structure, and the accretion by strings
which have a substantial amount of small-scale structure and hence
have a smaller translational velocity.  In the final section we
summarize the results and discuss future work.

We work in the context of an expanding, spatially flat
Friedmann-Robertson-Walker (FRW) Universe with scale factor $a(t)$
(normalized to $a(t_0) =1$ at the present time $t_0$).  The associated
redshift is $z(t)$. Newton's constant is denoted by $G$.

\chapter{Cosmic String Review}

Currently, two classes of models are receiving special attention as
possible theories for the origin of structure in the
Universe$^{9)}$. The first class is based on quantum fluctuations
produced during a hypothetical  period of exponential expansion
(inflation) in the very early Universe, and generically (but not
always) gives rise to a roughly scale invariant spectrum of primordial
adiabatic perturbations with random phases and a Gaussian probability
distribution.

The second class of models is based on topological defects which form
during a phase transition of matter in the very early Universe.  The
cosmic string theory$^{2)}$ belongs to this class.  Topological defect
models also give rise to a roughly scale invariant spectrum of
perturbations.  However, these fluctuations are not purely
adiabatic. More importantly, the models predict highly nonrandom
phases. 

Cosmic strings arise in a certain class of relativistic quantum field
theories which are believed to describe matter at very high
temperatures.$^{10)}$  They are formed during a phase transition from
a high temperature symmetric state to a low temperature state in which
the symmetry is broken. The formation mechanism is analogous to the
way in which defects form during crystallization of a metal and in
which vortex lines form during a temperature quench in superfluids and
superconductors.

The important point is that if the microscopic theory admits strings,
then a network of string inevitably forms during the phase
transition. Strings can have no ends and are therefore either
infinitely long or in the form of closed loops.  The presence of
strings simply reflects the fact that by causality there can be no
order over large distance scales. This causality argument in fact
implies that in a theory which admits strings, a network of strings
with mean separation of at most the Hubble radius will be present at
all times $t$ later than the phase transition.  In particular, strings
will be present at the time $t_{eq}$ when matter can start to accrete
onto seed perturbations.

The dynamics of cosmic strings is nontrivial.  The action of a single
string$^{11)}$ is the Nambu-Goto action.  The resulting equations lead
to relativistic velocities for string motion induced by bending of the
strings.  The Nambu-Goto action does not apply at the crossing points
of two strings.  Numerical simulations$^{12)}$ have shown that strings
do not pass through each-other but intercommute, i.e., break and
exchange ends.  In a cosmological setting, this leads to a mechanism
by which the long string network loses energy to string loops formed
when long strings intersect and which in turn slowly evaporate via
gravitational radiation.

As can be inferred by analytical arguments$^{4)}$ and was verified by
detailed numerical simulations,$^{7)}$ the evolution of the string
network approaches a scaling solution, a dynamical fixed point for
which the statistical properties of the string network are time
independent when all lengths are scaled to the Hubble radius
$t$. According to the scaling solution, on average a fixed number $N$
of long strings (strings with curvature radius greater than the Hubble
radius) cross any Hubble volume.  There is a remnant distribution of
loops which have been chopped off from the long string network at
earlier times.

The present simulations indicate that most of the energy of the string 
distribution is in long strings.  However, there are still substantial 
uncertainties in the details of the distribution.  String 
intercommutations induce kinks on the long strings which build up over 
time to give a substantial amount of small-scale structure.  Numerical 
simulations do not have enough resolution to track the small-scale 
structure.  This leads to an inherent uncertainty in the long string 
distribution since in the numerical analysis one is in essence coarse 
graining over the small-scale structure.  This leads to an even 
greater uncertainty in the loop distribution.

The uncertainty in the cosmic string distribution leads to 
uncertainties in the cosmic string structure formation scenario.  Long 
straight strings without small scale structure have no Newtonian 
gravitational potential.  If their$^{13)}$ transverse velocity is 
$v_s$, they induce a velocity perturbation of magnitude$^{14)}$
$$
\delta v = 4 \pi G \mu v_s \gamma (v_s) \>\>\>\,\,\, \gamma (v_s) = (1
- v_s^2)^{-1/2} \eqno\eq
$$
towards the plane behind the string, where $\mu$ is the mass per unit
length of the string. This gives rise to planar overdensities, the
so-called ``wakes."

Strings with a substantial amount of small-scale structure have a 
coarse grained tension $T$ less than $\mu$ 
and hence also produce, in addition to the above velocity perturbation, 
a Newtonian gravitational potential$^{15)}$
$$
h_{00} (r) = 4 G (\mu - T) \ln {r\over r_0} \eqno\eq
$$
where $r$ is the distance from the string and $r_0$ the string width.  
Since $T < \mu$ leads to sub-relativistic velocities for the string, 
the accretion pattern of such strings is typically more filamentary 
than planar.

In principle, the string network will induce inhomogeneities from the 
time it is formed.  However, for $t < t_{eq}$ the perturbations only  
grow logarithmically.  Velocity fluctuations in fact decay.  The
large-scale structure predicted by the string model is thus determined
by strings present at times $t \geq t_{eq}$.  Perturbations due to
strings at $t_{eq}$ are the most numerous and have the longest time to
grow.  They will hence determine the distinguished scale of structure
in the Universe$^{16,17)}$.

To study the formation of galaxies and galaxy clusters, it is crucial 
to understand the nonlinear evolution of the wakes and filaments.  
There has been little work on this subject.  In a pioneering paper, 
Rees$^{18)}$ discussed the evolution of baryons in a cosmic string 
wake.  Hara and Miyoshi$^{19)}$ performed an analytical and numerical 
analysis of perfectly planar wake formation including baryons.  Wake 
formation was studied by means of the Zel'dovich approximation for 
both cold and hot dark matter$^{5)}$.  This method only keeps track of 
the dark matter.  The Zel'dovich approximation was applied to the 
formation of string filaments in Refs. 20 and 21 for cold and hot dark 
matter respectively (see also Refs. 22 - 24).

In this paper we present the first results of a numerical study of the 
nonlinear dynamics of both dark matter and baryons in cosmic string 
wakes and filaments.  We use a recently developed cosmological hydro 
code$^{1)}$ which utilizes the PPM method to solve the hydrodynamical 
equations and a PIC technique to follow the dark matter particles.  
One of our major goals is to determine how significantly the internal 
structure of a wake or filament depends on the amount of small-scale 
structure on the string.

In order to be able to compare our simulations to some analytical 
results, we briefly review the analysis of wake formation using the 
Zel'dovich approximation$^{25)}$.  Quantities such as the overall 
thickness of the dark matter wake should be able to be reproduced with 
reasonable accuracy using this approximation.

The Zel'dovich approximation is based on writing the physical height 
$h$ of a dark matter particle above the wake in terms of a comoving 
perturbation $\psi$:
$$
h (q, t) = a (t) (q - \psi (q,t) ) \, , \eqno\eq
$$
$q$ being the initial comoving coordinate.  The basic equations used 
in deriving the equation of motion for $\psi$ are the Newtonian
gravitational force equation
$$
\ddot h = - {\partial\over{\partial h}} \Phi \, , \eqno\eq
$$
the Poisson equation for the gravitational potential $\Phi$
$$
{\partial^2\over{\partial h^2}} \Phi = 4 \pi G \rho \, , \eqno\eq
$$
and the mass conservation equation
$$
\rho (h,t) d^3 h = a^3 (t) \rho_0 (t) d^3 q \eqno\eq
$$
which relates the physical energy density $\rho (h,t)$ to the 
background density $\rho_0 (t)$.  After linearizing in $\psi$, the 
resulting equation for $\psi$ becomes$^{5)}$ (in the absence of any
gravitational line source on the string)
$$
\ddot \psi + 2 {\dot a\over a} \dot \psi + 3  {\ddot a\over a} \psi = 
0 \, . \eqno\eq
$$
The initial conditions at the time $t_i$ when the string is passing by 
are given by the velocity perturbation of Eq. 2.1, i.e.,
$$
\psi (q,t_i) = 0 \, , \>\>\> \dot \psi (q, t_i) = a (t_i)^{-1} 4 \pi G 
\mu v_s \gamma(v_s) \, . \eqno\eq
$$

The solution to (2.7) describes how the dark matter particles which are 
initially moving away from the wake with the Hubble flow eventually
get gravitationally bound to the wake.  The Zel'dovich approximation
gives a good description of the particle motion until the point of
``turn-around" when $\dot h = 0$. It follows from (2.3) that at the
time of turn-around $\psi = {1\over 2} q$.  The value of $q$ for which
particles are turning around at time $t$ for perturbations established
at a time $t_i$ is denoted by $q_{nl} (t_i , t)$.  The analysis of
Ref. 5 shows that
$$
q_{nl} (t_i , t) = {24 \pi\over 5} G \mu v_s \gamma (v_s) z 
(t_i)^{1/2} z (t)^{-1} t_0 \, . \eqno\eq
$$
The corresponding physical height above the center of the wake is
$$
h_{nl} (t_i, t) = {1\over 2} q_{nl} (t_i , t) z (t)^{-1} = {12 
\pi\over 5} G \mu v_s \gamma (v_s) z (t_i)^{1/2} z(t)^{-2} t_0 \, . 
\eqno\eq
$$
Once dark matter particles turn around, they will virialize$^{26)}$ at 
a distance of about ${1\over 2} h_{nl} (t_i , t)$ and will remain at 
this height: they have decoupled from the Hubble expansion.

For a simulation which starts with a velocity perturbation induced by 
a string at time $t_i$ and ends at time $t$, we expect the width of 
the dark matter distribution to be consistent with the value given above.

\chapter{Methods}

In order to study the nonlinear evolution of baryons and dark matter 
in cosmic string-induced wakes and filaments, we have performed 
numerical simulations making use of a new Eulerian PPM/PIC code for 
cosmological hydrodynamics$^{1)}$.  The construction and testing of 
the code are described in detail in Ref. 1.  Here we give a brief 
summary of the techniques used. 

The code follows the evolution of baryons and noninteracting dark 
matter in an expanding Universe.  The effects of strings are put in as 
external velocity or gravitational potential fluctuations.  The code 
does not take radiation into account, and hence cannot be used to
evolve the baryons before $t_{rec}$, the time of recombination.
Cooling of baryons is also not included in this code. The baryonic
fluid is treated as a single fluid of hydrogen. This approximation is
a reasonable one if we are interested in understanding shock behavior
and in the initial stages of nonlinear evolution. However, it does not
allow us to study the evolution and hydrodynamically induced
fragmentation of high density peaks.

There are two classes of simulation methods used to study cosmological
hydrodynamics numerically.  The first approach is grid-based
(Eulerian).  The hydrodynamical equations  are discretized and solved
on a fixed comoving grid.  The second  approach is smoothed particle
hydrodynamics (SPH) in which the fluid is  treated as a set of
particles statistically representing the fluid, with interactions
which follow from the hydrodynamic equations.  The Eulerian approach
has as its advantage good shock resolution, whereas the advantage of
SPH is the better resolution of high density regions. We chose the
grid-based Eulerian approach since we expect shocks to be very
important for cosmic string-induced structure formation and since we
are at the moment less interested in resolving the highest density 
peaks.

\section{Equations}

The baryonic fluid equations are obtained by setting the covariant 
divergence of the energy-momentum tensor equal to zero.  In the limit 
of non-relativistic velocities and pressure much less than the rest 
mass density of the fluid, the equations in comoving coordinates are
$$
\dot \rho + {\mathop {\bf \nabla}} \cdot (\rho {\bf v}) = 0 \eqno\eq
$$
$$
(\rho v_i)^\cdot + {\mathop {\bf \nabla}} \cdot (\rho v_i {\bf v} + 
{\bf e}_i p) = - 2 {\dot a\over a} \rho v_i - {\rho\over a^3} 
{\mathop {\bf \nabla}} \phi \eqno\eq
$$
$$
(\rho E)^\cdot + {\mathop {\bf \nabla}} \cdot (\rho E + p) {\bf v} = - 4 
{\dot a \over a}  \rho E - {\rho\over{a^3}} {\bf v} 
{\mathop {\bf \nabla}} \cdot \phi \, . \eqno\eq
$$
These are the equations$^{26)}$ which describe an inviscid fluid with 
no shear or stress terms.  In the above, $\rho$ is the matter density,
$p$ is pressure, $E$ is the total energy which can be written as
$$
E = {1\over 2} v^2 + u \, , \eqno\eq
$$
with $u$ denoting the internal energy; $v$ is the comoving peculiar
velocity and $\phi$ is the gravitational potential.  The variables
have been chosen such that the differential operators on the left hand
side of (3.1 - 3.3) have the same form as in the Euler equations in a
nonexpanding space (see e.g. Ref. 26).  They are related to physical
variables (with subscript $p$) via
$$
\rho = a^3 \rho_p , \, p=ap_p , \, a^2 u = u_p, \, a^2 T = T_p, \, a^2 
E = E_p, \, \phi = a \phi_p + {a^2\over 2} \ddot a x_p \, . \eqno\eq
$$
We also assume that the fluid is adiabatic
$$
p = c\rho T \, , \eqno\eq
$$
where $c$ is the gas constant, and obeys an ideal gas equation of state
$$
\rho u (\gamma - 1) = p \, . \eqno\eq
$$

Noninteracting dark matter constitutes the second component of our 
system.  Although present analyses indicate that the cosmic string 
model is in better agreement with the observed power spectrum of 
structures in the Universe if the dark matter is hot$^{5, 27)}$, we 
here for simplicity consider cold dark matter, collisionless particles 
with negligible thermal velocity dispersion\foot{Since we will
primarily study wakes which are formed at late times, hot dark matter
will already be cold at the relevant times, and hence will behave
similarly.}.  In this case, the equation of motion for dark matter
particles is the collisionless Boltzmann equation (Vlasov
equation). For particles in Lagrangian coordinates, the equation is:
$$
{\dot {\bf v}} + 2 {\dot a\over a} {\bf v}  =  {1\over{a^3}} 
{{\bf \nabla} \phi}
$$
$$  
{\dot {\bf x}}  = {\bf v} \, .  \eqno\eq
$$
The only forces are due to gradients in the gravitational potential.

In the Newtonian approximation to the Einstein equations, the 
gravitational potential $\phi$ is determined via the Poisson equation 
from the matter density $\rho$ of both baryonic and dark 
matter:
$$
\nabla^2 \phi = 4 \pi G (\rho - \bar \rho) \eqno\eq
$$
where $\bar \rho$ is the spatial average of $\rho$.

The effects of strings enter as initial conditions in our analysis.  
At the beginning of the simulation, we take the distribution of 
baryons and dark matter to be unperturbed.  The baryonic matter 
density is taken to be 5\% of the total matter density.  In order to 
justify neglecting the radiation pressure, we take our initial time to 
be after $t_{rec}$, i.e., at a redshift $z_{in} < 1200$.  The 
expansion rate of the Universe can then be taken to be
$$
a (t) = \left( {t\over t_0} \right)^{2/3} \, . \eqno\eq
$$

The strings produce velocity and gravitational potential fluctuations.  
We consider a straight string at $z = 0$ with tangent vector 
${\bf e}_x$ moving with velocity $v_s$ in $y$ direction. Such a string
will produce a velocity perturbation of magnitude given by Eq. 2.1
towards the $x-y$ plane.  The velocity perturbation for a fixed value
of $y$ sets in once the string passes this value of $y$ and then
propagates with the speed of light in the $z$ direction.

For strings without small-scale structure, there is no initial 
gravitational potential perturbation.  The effects of strings with 
small scale structure can be modelled by giving the string an 
effective Newtonian mass per unit length $\mu_{eff}$ of magnitude
$$
\mu_{eff} = \mu - T \eqno\eq
$$
(see Eq. 2.2), which via the Poisson equation induces a Newtonian 
gravitational potential fluctuation.  

\section{Numerical Techniques and Implementation}

The key issue in selecting a numerical technique for solving the
hydrodynamical equations is accurate resolution of non-linear
effects. Of particular importance is good shock resolution since
shocks are expected to play a crucial role in cosmological flows, in
particular in the case of cosmic string wakes. 

PPM (piecewise parabolic method)$^{28)}$ is a technique which has been
well tested as an accurate method for treating hydrodynamical flows
with discontinuities. Hence, we have chosen the PPM algorithm to
evolve the baryonic fluid. Since PPM is grid-based, it is most natural
to use a grid-based method to evolve the dark matter distribution. We
use the PIC (particle in cell) method$^{29)}$, an extensively tested
scheme which combines particle and grid methods to evolve the dark
matter in our system (This method is more commonly referred to as the
particle-mesh (PM) method, but we use the acronym PIC to avoid
confusion of PM with PPM). We have combined the PIC and PPM codes to
form a cosmological hydro code which simultaneously evolves our two
fluid system consisting of collisionless dark matter and collisional
baryons.

PPM is a higher order Godunov method for integrating partial
differential equations. Our code is based on a code originally
developed for non-linear astrophysical problems$^{30)}$.

The Godunov method is a finite volume scheme. The equations of motion
are considered in their integral form. Thus, the problem of
calculating spatial gradient terms becomes a problem of determining
fluxes. The advantage of this procedure in that mass, energy and
momentum are exactly conserved in the absence of source terms such as
the expansion of the Universe.

The simulation volume is divided into a set of cells, for each of
which the average values of the fluid variables are kept track. To
find an approximate solution of the integrated Euler equations, one
needs to determine the fluxes across the cell boundaries. For
instance, the integrated continuity equation
$$
\int d^3x \partial_t \rho + \int d^3 x \nabla \cdot (\rho {\bf v}) = 0
\eqno\eq 
$$
(integration over one cell) becomes
$$
\partial_t \bar \rho + \sum_{\rm sides} \rho {\bf v} \cdot {\bf S} =
0,\eqno\eq
$$
where $\bar \rho$ is the total density in the cell and ${\bf S}$ is
the normal vector to the side of the cell. Obviously, a prescription
is needed in order to be able to compute the fluxes based on knowing
only the average values of the fluid variables in the cells.

In the Godunov method, the fluxes are computed in two steps. First,
profiles of the fluid variables in each cell are constructed based on
the average values in the cell and its neighbors. In the second step,
the Riemann shock tube problem is solved at the cell interfaces,
giving a set of nonlinear discontinuities in the fluid variables
propagating away from the interface with characteristic
velocities. These propagating discontinuities give the fluxes from and
to each cell, which are used in the final step of the algorithm to
update the fluid variables. For a discussion of the Riemann shock tube
problem (which is in essence a way of solving the integrated form of
the Euler equations across a discontinuity) see Refs. 1 and 31. 

The advantage of Godunov methods is that the non-linearities in the
evolution equations are incorporated directly in the differencing
scheme via the solution of the Riemann shock tube problem. Linear
schemes for calculating fluxes are unable to simultaneously well
reproduce both the width of the discontinuity and the amplitude of the
waves travelling away from it. Linear schemes may also spuriously
allow sound waves to propagate supersonically. Both of these problems
are avoided by using the Godunov method.

PPM introduces a number of changes$^{28)}$ to achieve higher order
resolution in the Godunov method. Most importantly, instead of using
profiles of the fluid variables which are constant across any cell,
interpolating parabolae are employed. This gives better spatial
resolution and allows a more precise determination of the initial data
for the Riemann shock tube problem solving routine. In order to damp
spurious oscillations at shocks, the parabolae are flattened out near
shocks. As a consequence, a much smaller artificial numerical
viscocity is required in order to damp the oscillations. 

The initial PPM code$^{30)}$ was written for hydrodynamics in
non-expanding space. The inclusion of dynamical gravity necessitates
two main changes. First, the gravitational potential is introduced. It
is computed at each time step and at each grid point by solving the
Poisson equation (3.9) (with $\rho$ being the sum of baryon and dark
matter mass density) by means of a standard FFT scheme. The second
major change is, as can be seen from Eqs. 3.2 and 3.3, the appearance
of additional source terms in the fluid equations. These terms are
local and hence do not effect the computation of gradients. They are
incorporated into our code using standard operator splitting methods
while keeping the code accurate to second order.

Since baryonic and dark matter interact with each other only
gravitationally, it is straightforward to combine the PPM and PIC
codes. Some nontrivial issues concerning this combination of codes are
discussed in Ref. 1.

The code was implemented on a MasPar MP-2 at the Goddard Space Flight
Center. It was tested extensively including  and without the expansion
of the Universe. For small sinusoidal fluctuations, the numerical
results were tested against the analytical predictions of linear
theory (Jeans test). Over a time interval of 40 Hubble expansion
times, the relative error between the numerical and analytical values
for the amplitude was less than 0.5\%, given an initial perturbation
amplitude equal to $10^{-3}$\% of the background (see Ref. 1 for
further discussions of this and other tests). Tests also showed that
the code is able to resolve features of size corresponding to 2 or 3
grid spacings. On scales of 32 grid spacings or larger, the agreement
between analytical and numerical results is better than 99\% over a
time period of 10 Hubble expansion times. The code conserves energy to
better than 98\% for a simulation which starts at redshift 100 and
ends at redshift 8.

\chapter{Results}

As a first application of our new hydro code, we studied the accretion
of baryons and cold dark matter by individual cosmic strings. We
considered two extreme possibilities: a rapidly moving string without
small-scale structure and thus without a local Newtonian gravitational
potential, and a slowly moving string with small-scale structure and
induced gravitational force towards the string. In the first case, the
structures which form are planar wakes, in the second case the
accretion leads to more filamentary structures.

\section{Planar Accretion}

As an idealization of the accretion pattern produced by a cosmic
string with tangent vector pointing along the $x$ coordinate axis and
moving with velocity $v_s \sim 1$ in $y$ direction, we first study
planar accretion.

The initial values of the fluid variables and the dark matter particle
distribution were taken to be homogeneous. At a time $t_i$ (with
corresponding redshift $z_i$), we use a velocity perturbation of
physical amplitude
$$
\delta v = 4 \pi G \mu v_s \gamma(v_s) \eqno\eq
$$
towards the plane $x = y = 0$ as initial condition. Note that the
velocity is independent of the distance from the plane\foot{The
velocity goes to zero at a distance comparable to the Hubble radius at
$t_i$, which is taken to be much larger than our simulation
volume$^{32)}$.}. The velocity perturbation effects the baryonic fluid
and the dark matter equally, at least for times greater than the time
of recombination after which Jeans damping of the baryon perturbations
is negligible on the cosmological scales of interest). We made tests
to ascertain that these initial conditions are a good approximation to
the actual velocity perturbations produced by the string in a full
axially symmetric simulation.

Since the initial velocity field is independent of the distance from
the $x-y$ plane, the perturbation is in marked contrast to the
velocity field in a Zel'dovich$^{23)}$ pancake resulting from
clustering in a model with adiabatic perturbations, in which case the
velocity increases rapidly towards the central plane. This is due to
the nonvanishing initial gravitational potential. As we shall see
later, this difference in the initial velocity field may have
important consequences for biasing.

In Figures 1 - 16 we illustrate some of the main results of our planar
collapse analyses. The simulations were performed for the values $v_s
= 0.5$ and $G \mu = 10^{-6}$, the value of the mass per unit length of
cosmic strings determined from analytical estimates of gravitational
clustering$^{5)}$ and CMB anisotropy generation$^{3)}$. It is,
however, trivial to rescale our results to different values of $G \mu$
and $v_s$, at least for the dark matter distribution.

Figures 1 - 6 show the density of baryons and dark matter at different
redshifts along a cross section through the string wake. As in all of
our present simulations, the average baryon density was 5\% of
critical density whereas $\Omega_{CDM} = 0.95$. The initial redshift
was $z_i = 100$. We see that baryons and dark matter evolve quite
differently. Whereas the baryons streaming in from both sides of the
wake collide in the center, form shocks, heat the gas and give off
most of their kinetic energy, the collisionless dark matter particles
stream through the central plane and subsequently oscillate about it,
thus forming density peaks at the various turn-around surfaces.

In order to check the results of the MasPar simulations and in order
to obtain higher resolution, a one-dimensional PIC code was
written. This allows a much finer sampling of the dark matter
distribution along a cross-section through the wake. Figure 7 depicts
the phase space diagram of dark matter particles obtained using this
code. The simulation began at a redshift $z_i = 100$. The figure shows
the result at $z = 2$. Clearly visible are the various turn-around
distances which correspond to the density peaks in the dark matter
distribution of Figures 1 - 6. Note that the initial turn-around
surface (which corresponds to break-away from the Hubble flow) does
not give rise to a density peak. Hence, the wake thickness as
determined from Figure 7 (the width of the region between the
outermost particle turnaround radii) is only about 0.14 of
the ``thickness" $2 q_{nl}(z)$ calculated using the Zel'dovich
approximation. This factor is consistent with the work of Filmore and
Goldreich$^{34)}$ on self-similar clustering of a dark matter
distribution with planar symmetry.

In Figure 8 we plot the time dependence of the particle turn-around
radius. The numerical results (cyan crosses) are compared with the
analytical prediction from Eq. 2.9 that $q_{nl} \propto t^{2/3}$. The
agreement over the entire period of the run, beginning at redshift
$z_i = 10000$ and ending at $z \sim 1$ is very good. The predicted width
of the nonlinear region today is $w \simeq 0.7 Mpc$, consistent with
the value given by Eq. 2.9.

The most interesting result of our numerical study is the explicit
demonstration that planar collapse onto cosmic string wakes will lead
to a relative baryon enhancement in the central region of the wake. As
can be read off from Figure 9, the enhancement factor is 2.4. This
ratio is time-independent once the accretion pattern achieves its
self-similar form. As demonstrated in Figure 8, the approach to
self-similarity is very rapid on the time scale of a cosmological
simulation. The thickness of the region of relative baryon enhancement
is, when extrapolated using the self-similar solution, predicted to be
about 0.3Mpc. This result may help explain - in the context of the
cosmic string model - the recently detected high baryon abundance in
some rich clusters$^{34)}$. However, in order to make any definite
statements, we must still understand the formation of galaxy clusters
in the cosmic string theory. This is one of the goals of our future
research. If, as suggested in Ref. 35, clusters are associated with
the crossing sites of three wakes, the baryon overdensity factor in a
cluster core would be larger than the corresponding factor for the
central region of an individual wake. 

We also have detailed information about the temperature profile
through a wake. Figures 10 - 15 show the temperature of the baryonic
gas along a cross-section of the wake at different times. The
temperature peaks are at the locations of the strong shocks where the
infalling stream of baryons hits the distribution of baryons which
have fallen in previously and lost their kinetic energy to shocks.
The temperature at the shock positions is determined by the conversion
of kinetic to thermal energy. For an ideal cosmic string wake studied
here, the pre-shock flow of baryons is a cool and coherent
flow. Hence, the resulting post-shock temperature $T_s$ is low:
$$
T_s \sim m_H (\delta v)^2 \sim m_H v_s^2 (4 \pi G \mu)^2 \sim 10^2 K,
\eqno\eq
$$
where $m_H$ is the mass of a hydrogen atom.

Note that for wakes formed earlier than $z \sim 800$, the initial
thermal velocities of the baryons dominate over the string-induced
velocity perturbation. This leads to an initial diffusion of the
baryons over a distance larger than the width of the dark matter
density enhancement (see Figure 16).

\section{Moving Newtonian Line Source}

There are two main motivations to study clustering of dark and
baryonic matter induced by a slow moving string with a substantial
amount of small-scale structure, modelled as a moving Newtonian line
source. First, there are indications that over time a substantial
amount of small-scale structure builds up on the long string network,
and that hence this situation might well be realized for cosmic
strings resulting from grand unified phase transitions. A second
motivation for performing these simulations is in an attempt to
identify distinctive predictions of the string model in the regime of
nonlinear gravitational clustering. A moving Newtonian line source
would give rise to planes and filaments which would have similar
clustering properties as those objects seen to emerge in N-body
simulations of large-scale structure formation in models based on
adiabatic perturbations with a scale-invariant spectrum. It is
interesting to use the same numerical code to analyze and compare the
baryon and dark matter distribution resulting from nonlinear
clustering for string wakes (no initial Newtonian potential, only
deficit angle) and for initial perturbations with a Newtonian
source. In particular, it is of interest to study any relative baryon
enhancement in these models.

Figures 17 - 19 show results from simulations of clustering induced by
a slow moving cosmic string. The string tension was $T = 0.95 \mu_0$,
with $\mu_0$ being the non-renormalized mass per unit length. The
renormalized mass per unit length $\mu$ is related to $\mu_0$ and $T$
via$^{8, 36)}$
$$
T \mu = \mu_0^2 \, .\eqno\eq
$$
The initial string velocity was taken to be $v_s = 0.0005$. With these
parameter values, the Newtonian gravity of the string has a much
larger effect than the deficit angle. This can be seen by computing
the relative velocity $u$ in the $z$-direction developed by two particles
after a string has passed between them$^{20-22)}$:
$$
u = 8 \pi G \mu v_s \gamma(v_s) + {{4 \pi G (\mu - T)} \over {v_s
\gamma(v_s)}},
\eqno\eq
$$
where the first term is due to the deficit angle and the second due to
the Newtonian force. For the value $G \mu = 10^{-6}$ used in the
simulations, the term due to Newtonian gravity dominates by a factor
of $2 \times 10^5$.

As shown in Figures 17 and 18 at $z \sim 8$, taken from a simulation
with $z_i = 100$, the distribution of baryons and dark matter
perpendicular to the plane spanned by the tangent vector of the string
and its velocity vector is completely different from the corresponding
distribution for strings without small-scale structure (see
Fig. 1). Now, the baryons no longer remain confined to the central
region of the wake. There is a build-up of pressure in the vicinity of
this plane which imparts an outward velocity to the baryons: the
baryons are shock heated and expelled from the central
regions. Instead of a baryon overdensity, there is a relative baryon
deficit at $z \sim 80$.

From Figure 17 it is also manifest that the accretion pattern is no
longer planar, but rather filamentary. There are substantial
velocities in the $x$ direction towards the instantaneous location of
the string.

The temperature distribution of the gas along the $z$ axis is shown in 
Figure 19 (also for redshift 8). Since the baryon velocity at the
shock location is much larger than in the case of the wakes studied in
the previous section, the post-shock temperature is significantly
higher. From Eq. (4.4) it follows that for our values of $T/\mu$ and
$v_s$, the velocity flow $u$ induced by the filament is about a factor
$10^2$ larger than the velocity $\delta v$ obtained for wakes in the
previous simulations. Hence, by (4.2), the post-shock temperature is
expected to be higher than $10^6$K, as is confirmed in Fig. 19.

\chapter{Discussion}

Making use of a new three-dimensional cosmological hydro code, we have
simulated the clustering of baryons and cold dark matter induced by
long straight strings with and without small-scale structure. We have
studied the clustering induced by a single string. 

Strings with no small-scale structure give rise to planar wakes. We
have shown that the dark matter distribution is in excellent agreement
with the known self-similar solutions for planar accretion. In
particular, the Zel'dovich approximation yields a good estimate for
the thickness of the dark matter wake. For $G \mu = 10^{-6}$ and $v_s
= 0.5$, the thickness is about $0.7$ Mpc for perturbations generated
at $t_{eq}$, the time of equal matter and radiation.

We have shown that the baryon overdensity in wakes for strings without
small-scale structure is thinner than that of the dark matter, leading
to a relative enhancement of the baryon density in the center of the
wake. In our simulations with the above parameters, the enhancement
factor was about $2.4$, with a thickness of the baryon wakes of only
about $0.3$ Mpc.

For strings with a substantial amount of small-scale structure, the
baryon distribution is completely different. Instead of a baryon
density enhancement, there is a deficit in the center of the
structures. The difference is due to the fact that small-scale
structure on a string induces a Newtonian gravitational line source on
the string. The velocity of sound is now larger than the impulse due
to the Newtonian potential of the string. Hence baryons can more
easily thermalize, a high pressure builds up, creating a rapidly
outward moving shock.

The difference between string wakes and string filaments is very
pronounced when considering the post-shock baryon temperatures. For
wakes, the cold coherent flow leads to temperatures of only about
$10^2$K, whereas the large velocities induced by the Newtonian
gravitational line source on the string filaments lead to very high
temperatures. For the values of $T/\mu = 0.95$ and $v_s = 0.0005$ used
in our simulations, the post-shock temperature was about $10^6$K,
comparable to the temperature of clusters in inflation-based CDM
models$^{37, 38)}$. The difference in baryon temperatures between the
string wake and string filament models may lead to very different
ionization histories\foot{We thank Douglas Scott for a useful
discussion on this subject.}. Based on our simulations we conclude
that the temperature in string wakes is too low to lead to ionization
at high redshifts since no atomic cooling will occur, whereas in
string filaments the baryons are sufficiently hot to lead to
substantial cooling, emission of energetic photons, and subsequent
ionization. These issues will be explored in future work.

We can compare our results on the segregation of baryons and cold dark
matter with recent results of U.-L. Pen$^{38)}$, who has studied local
properties of gas in rich clusters of galaxies in a theory with
Gaussian adiabatic perturbations with a scale-invariant spectrum by
means of numerical simulations using a new adaptive mesh hydro
code$^{39)}$. This code - apart from the adaptive moving mesh - is
based on similar numerical techniques as our code. The results of
these simulations show a marked baryon deficit in the cluster
centers. Note, however, that neither our code nor the code of Pen
include cooling, which might change the results. 

However, by comparing our results for strings with and without
small-scale structure, we have identified a further distinguishing
characteristic of the cosmic string wake model: there will be a baryon
enhancement in the central region of the wakes. If, as proposed by
Hara et al.$^{35)}$, rich clusters of galaxies are identified with the
crossing sites of three wakes, the baryon enhancement factor in
clusters would be about three times that in a wake, i.e. about
7. Thus, a cosmic string model may be able to explain in a natural way
the observed high baryon fraction in clusters such as COMA$^{34)}$ in
the context of a spatially flat Universe. 

A further distinguishing feature of the string model which has been
confirmed by our simulations concerns the geometry of the nonlinear
density perturbations produced. The most numerous and thickest string
wakes were generated by strings at $t_{eq}$. Their comoving length $l$
and width $w$ are given by 
$$
w \propto l \propto \xi t_{eq} z(t_{eq}) v_s \gamma(v_s) \propto \xi
\times 40 h^{-1} {\rm Mpc}, \eqno\eq
$$
where $\xi$ is a constant of order unity which sets the curvature
radius of the long strings relative to the Hubble radius. The thickness
of these planar structures is about $0.7$ Mpc. These predictions are
in encouraging agreement with recent observations$^{40)}$ which
indicate that the super-large-scale structure of the Universe is
dominated by planar structures of dimension comparable to $50 h^{-1}$
Mpc, and that these walls are indeed very thin$^{41)}$. 

\medskip
\centerline{\bf Acknowledgements}

The work of A.S. was supported in part by a NASA Graduate Student
Researcher award, and by the U.K. PPARC. The work of R.B. was
supported in part by the US Department of Energy under Grant
DE-FG0291ER40688, Task A. One of us (R.B.) thanks Douglas Scott and
Ue-Li Pen for fruitful discussions, and the Physics and Astronomy
Department of the University of British Columbia for hospitality.

\medskip
\REF\SFOM{A. Sornborger, B. Fryxell, K. Olson and P. MacNeice, "An
Eulerian PPM/PIC Code for Cosmological Hydrodynamics", Brown preprint
BROWN -HET-1000 (1995).}
\REF\CSreviews{For recent reviews see e.g. \nextline
A. Vilenkin and E.P.S. Shellard, {\it Cosmic strings and other
topological defects} (Cambridge Univ. Press, Cambridge,
1994);\nextline
M. Hindmarsh and T.W.B. Kibble, {\it Rept. Prog. Phys.} {\bf 58}, 477
(1995);\nextline
 R. Brandenberger, {\it Int. J. Mod. Phys.} {\bf A9}, 2117 (1994).}
\REF\CMBstrings{D. Bennett, A. Stebbins and F. Bouchet, {\it
Ap. J. (Lett.)} {\bf 399}, L5 (1992);\nextline
L. Perivolaropoulos, {\it Phys. Lett.} {\bf B298}, 305 (1993).}
\REF\ZelVil{Ya. B. Zel'dovich, {\it Mon. Not. R. Astr. Soc.} {\bf
192}, 663 (1980);\nextline
A. Vilenkin, {\it Phys. Rev. Lett.} {\bf 46}, 1169 (1981).}
\REF\PBS{L. Perivolaropoulos, R. Brandenberger and A. Stebbins, {\it
Phys. Rev.} {\bf D41}, 1764 (1990);\nextline
R. Brandenberger, L. Perivolaropoulos and A. Stebbins, {\it
Int. J. Mod. Phys.} {\bf A5}, 1633 (1990);\nextline 
R. Brandenberger, {\it Phys. Scripta} {\bf T36}, 114 (1991).}
\REF\StringCMB{B. Allen, R. Caldwell, E. P. S. Shellard, A. Stebbins
and S. Veeraraghavan, in {\it CMB Anisotropies Two Years After COBE},
ed. L.~Krauss (World Scientific, New York, 1993);\nextline
B. Allen, R. Caldwell, E. P. S. Shellard, A. Stebbins
and S. Veeraraghavan, ``Large Angular Scale CMB Anisotropy Induced
by Cosmic Strings'',  submitted to Phys. Rev. Lett. (May, 1996).}
\REF\CSsimuls{D. Bennett and F. Bouchet, {\it Phys. Rev. Lett.} {\bf
60}, 257 (1988);\nextline
B. Allen and E.P.S. Shellard, {\it Phys. Rev. Lett.} {\bf 64}, 119
(1990);\nextline
A. Albrecht and N. Turok, {\it Phys. Rev.} {\bf D40}, 973 (1989).}
\REF\Carter{B. Carter, {\it Phys. Rev.} {\bf D41}, 3869 (1990).}
\REF\RBcomp{For a recent comparative review see e.g.\nextline
R. Brandenberger, ``Formation of Structure in the Universe", Brown
preprint BROWN-HET-1006 (1995).}
\REF\Kibble{T.W.B. Kibble, {\it J. Phys.} {\bf A9}, 1387 (1976).}
\REF\Nielsen{H. Nielsen and P. Olesen, {\it Nucl. Phys.} {\bf B61}, 45
(1973).}
\REF\CScrossing{E.P.S. Shellard, {\it Nucl. Phys.} {\bf B283}, 624
(1987);\nextline 
R. Matzner, {\it Computers in Physics} {\bf 1}, 51 (1988);\nextline
K. Moriarty, E. Myers and C. Rebbi, {\it Phys. Lett.} {\bf 207B}, 411
(1988); \nextline 
E.P.S. Shellard and P. Ruback, {\it Phys. Lett.} {\bf 209B}, 262 (1988).}
\REF\Vildeficit{A. Vilenkin, {\it Phys. Rev.} {\bf D23}, 852 (1981).}
\REF\SilkVil{J. Silk and A. Vilenkin, {\it Phys. Rev. Lett.} {\bf 53},
1700 (1984).}
\REF\Vilbook{see e.g. A. Vilenkin and E.P.S. Shellard, `Cosmic strings
and other topological defects' (Cambridge Univ. Press, Cambridge,
1994).}
\REF\TVwakes{T. Vachaspati, {\it Phys. Rev. Lett.} {\bf 57}, 1655 (1986).}
\REF\SVBST{A. Stebbins, S. Veeraraghavan, R. Brandenberger, J. Silk
and N. Turok, {\it Ap. J.} {\bf 322}, 1 (1987).}
\REF\Rees{M. Rees, {\it Mon. Not. R. Astr. Soc.} {\bf 222}, 27p (1986).}
\REF\Hara{T. Hara and S. Miyoshi, {\it Prog. Theor. Phys.} {\bf 77},
1152 (1987);\nextline
T. Hara and S. Miyoshi, {\it Prog. Theor. Phys.} {\bf 84}, 867 (1990).}
\REF\Vollickone{D. Vollick, {\it Phys. Rev.} {\bf D45}, 1884 (1992).}
\REF\Vollicktwo{D. Vollick, {\it Ap. J.} {\bf 397}, 14 (1992).}
\REF\CSfilone{T. Vachaspati and A. Vilenkin, {\it Phys. Rev. Lett.}
{\bf 67}, 1057 (1991).}
\REF\CSfiltwo{T. Vachaspati, {\it Phys. Rev.} {\bf D45}, 3487 (1992).}
\REF\CSfilthree{A. Aguirre and R. Brandenberger, {\it
Int. J. Mod. Phys.} {\bf D4}, 711 (1995);\nextline
V. Zanchin, J.A.S. Lima and R. Brandenberger, ``Accretion of Hot and
Cold Dark Matter onto Cosmic String Filaments', Brown preprint
BROWN-HET-1049 (1996).}
\REF\Zel{Ya. B. Zel'dovich, {\it Astr. Astrophys.} {\bf 5}, 84
(1970).}
\REF\Peeblestwo{see e.g. P.J.E. Peebles, `The Large-Scale Structure of
the Universe' (Princeton Univ. Press, Princeton, 1980).}
\REF\astwo{A. Albrecht and A. Stebbins, {\it Phys. Rev. Lett.} {\bf
69}, 2615 (1992).}
\REF\Collela{P. Collela and P. Woodward, {\it J. Comp. Phys.} {\bf
54}, 174 (1984).}
\REF\Nbody{R. Hockney and J. Eastwood,  `Computer Simulations using Particles'
(Adam Hilger, 1988).}
\REF\Fryxell{B. Fryxell, E. Muller and D. Arnett, {\it Ap. J.} {\bf
367}, 619 (1991).}
\REF\Hirsch{C. Hirsch, `Numerical Computation of Internal and External
Flows' (John Wiley \& Sons, 1988).} 
\REF\Joao{J. Magueijo, {\it Phys. Rev.} {\bf D46}, 1368 (1992).}
\REF\Filmore{J. Filmore and P. Goldreich, {\it Ap. J.} {\bf 281}, 1
(1984).}
\REF\clusters{S. White, J. Navarro, A. Evrard and C. Frenk, {\it
Nature} {\bf 366}, 429 (1993).}
\REF\haratwo{T. Hara and S. Miyoshi, {\it Ap. J.} {412}, 22
(1993);\nextline
T. Hara, H. Yamamoto, P. M\"ah\"onen and S. Miyoshi, {\it Ap. J.} {\bf
432}, 31 (1994).}
\REF\vilninety{A. Vilenkin, {\it Phys. Rev.} {\bf D41}, 3038 (1990).}
\REF\numsim{see e.g. H. Kang, J. Ostriker, R. Cen, D. Ryu,
L. Hernquist, A. Evrard, G. Bryan and M. Norman, {\it Ap. J.} {\bf
430}, 83 (1994), and references quoted therein.}
\REF\penone{U.-L. Pen, ``The Local Properties of Gas in Individual
Rich Clusters of Galaxies", CFA preprint (1996).}
\REF\pentwo{U.-L. Pen, ``The Combined Cosmological Adiabatic
Hydrodynamics and N-body Code: MMHPM", CFA preprint (1996).}
\REF\dorosh{A. Doroshkevich, D. Tucker, A. Oemler, R. Kirshner,
H. Lin, S. Shechtman and S. Landy, ``Large- and Superlarge-Scale
Structure in the Las Campanas Redshift
Survey", TAC preprint 1995-030, subm. to
Mon. Not. R. astr. S. (1995).} 
\REF\CFA{V. de Lapparent, M. Geller and J. Huchra, {\it Ap. J.} {\bf
369}, 273 (1991).}
\refout
\vfill\eject
\centerline{\bf Figures}
\inserttwofigs[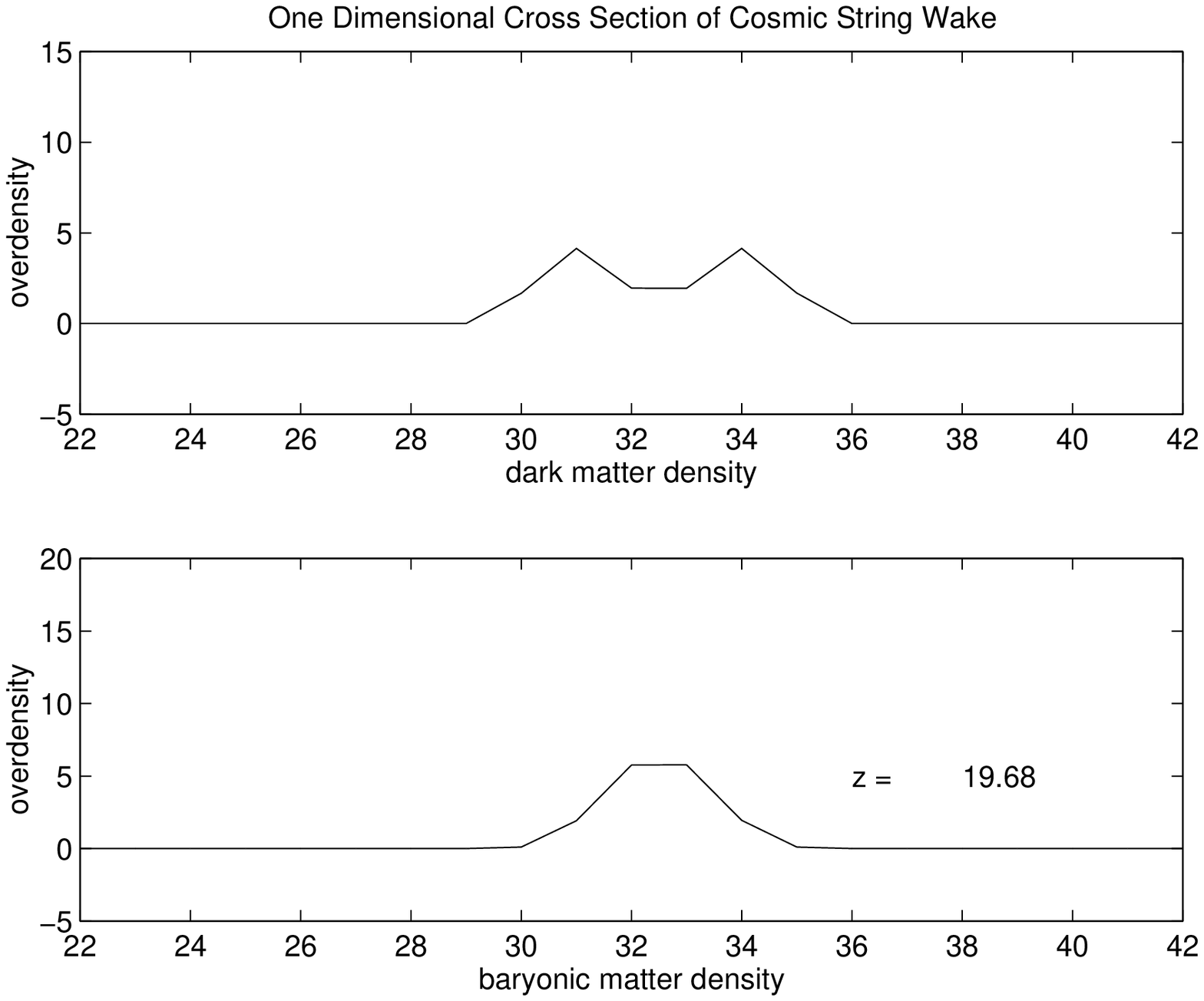,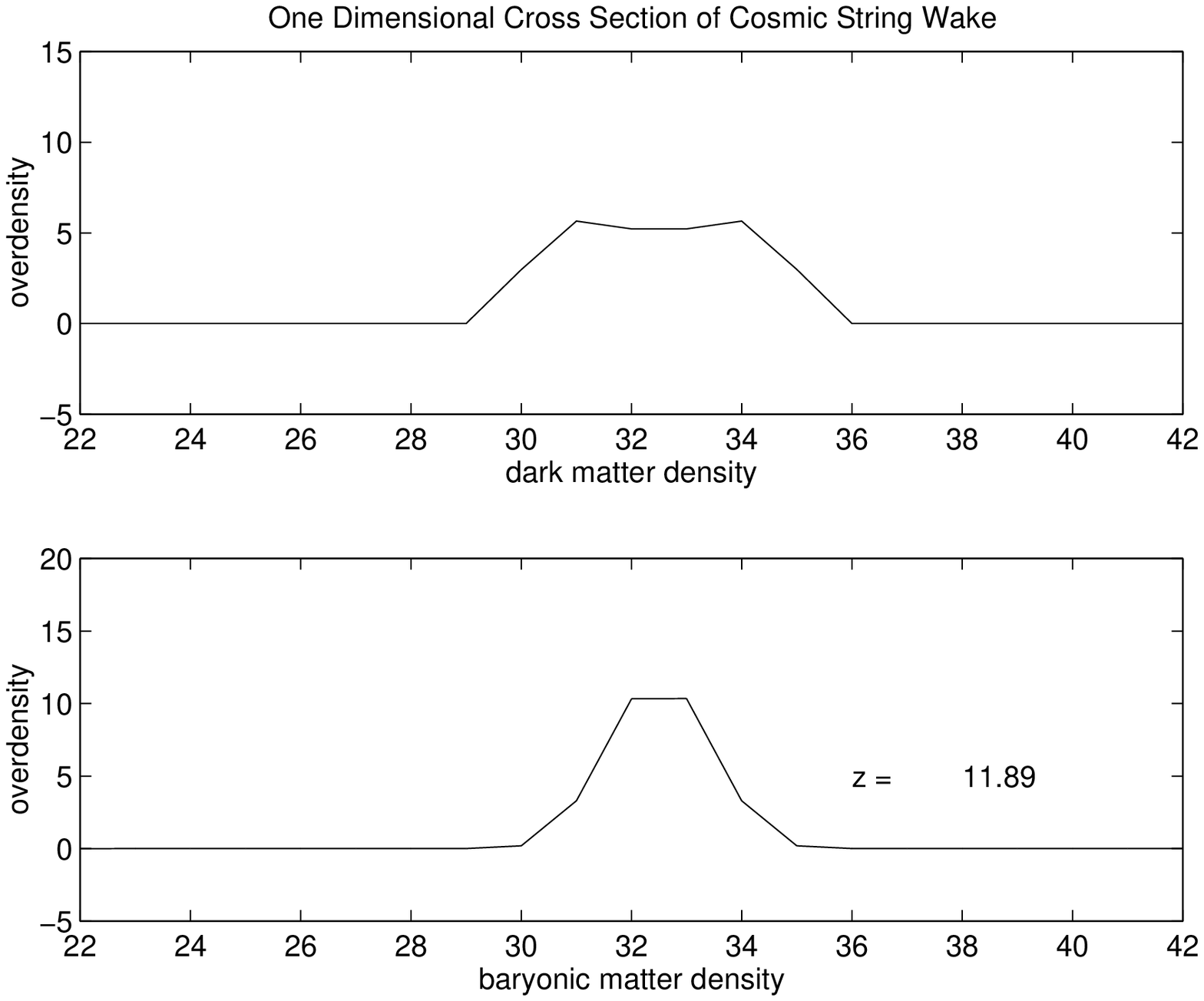]{}{}
\inserttwofigs[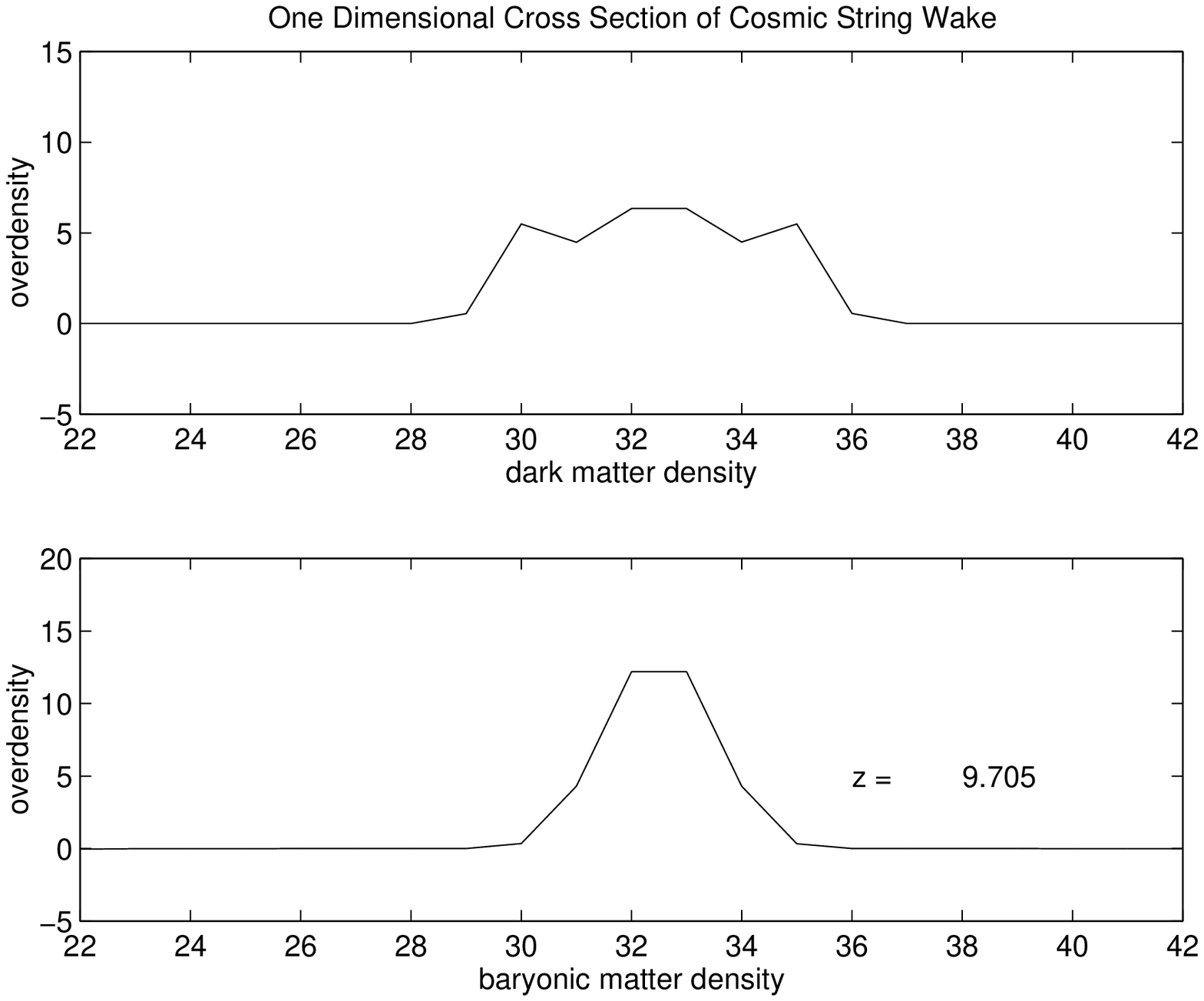,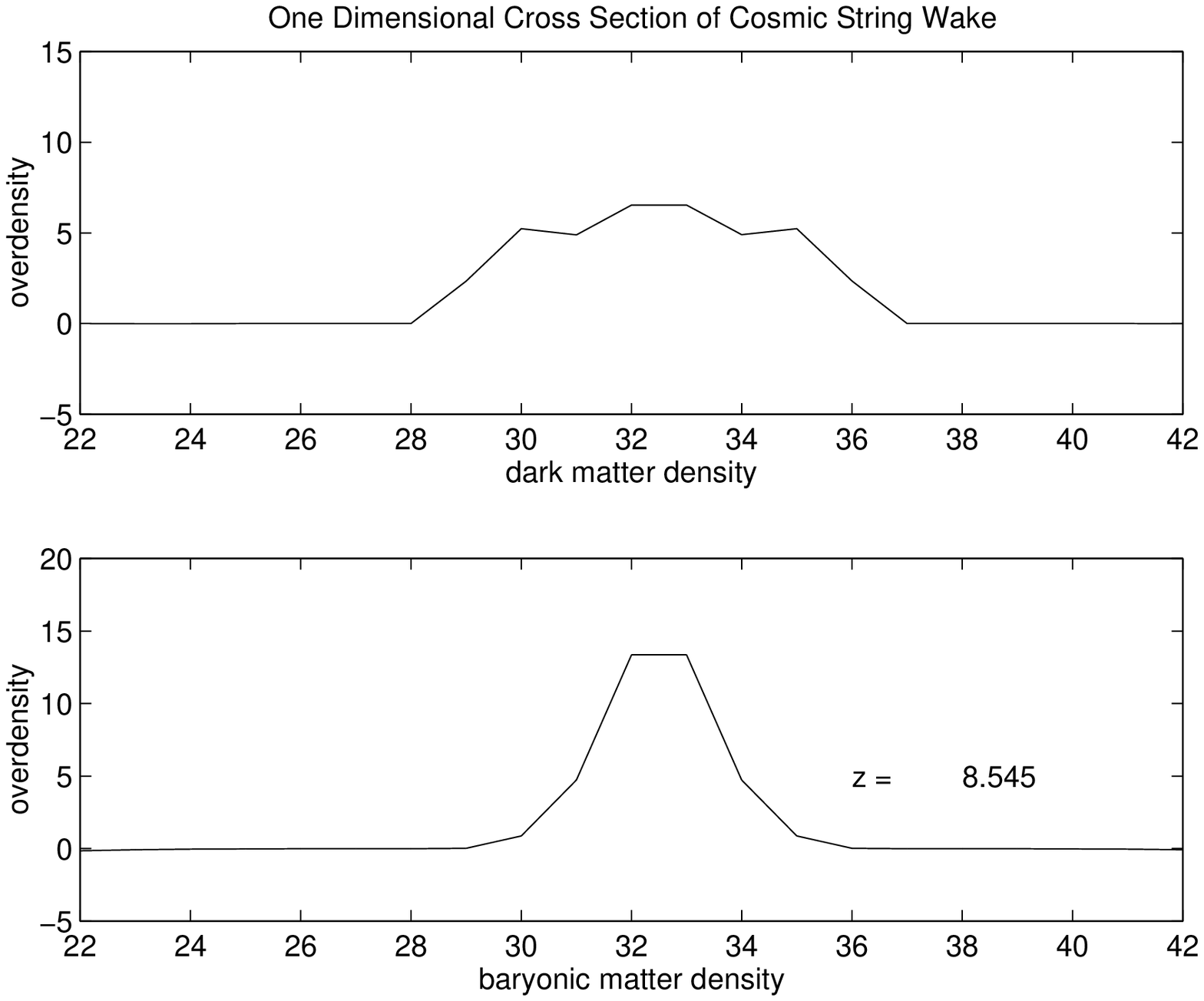]{}{}
\inserttwofigs[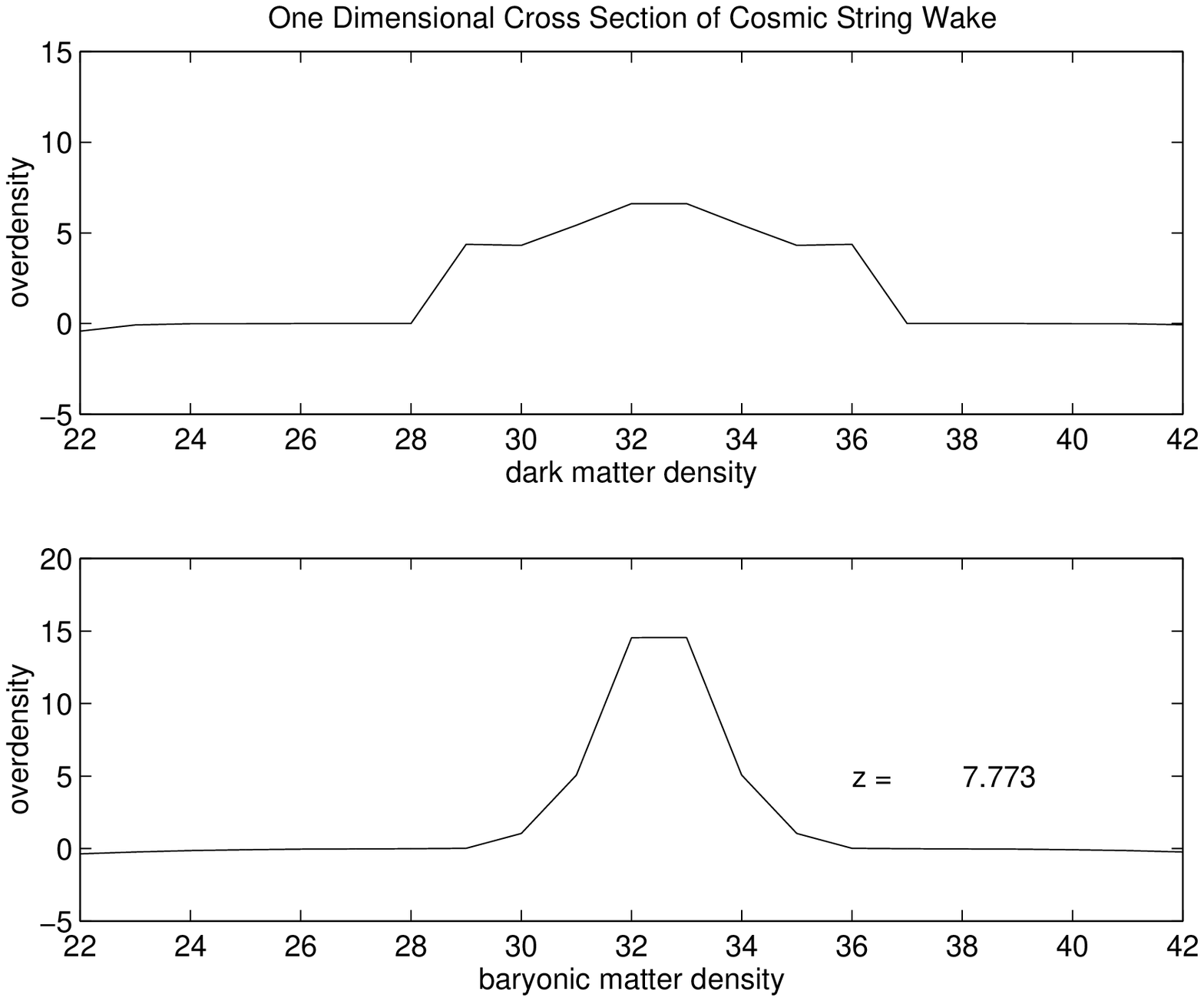,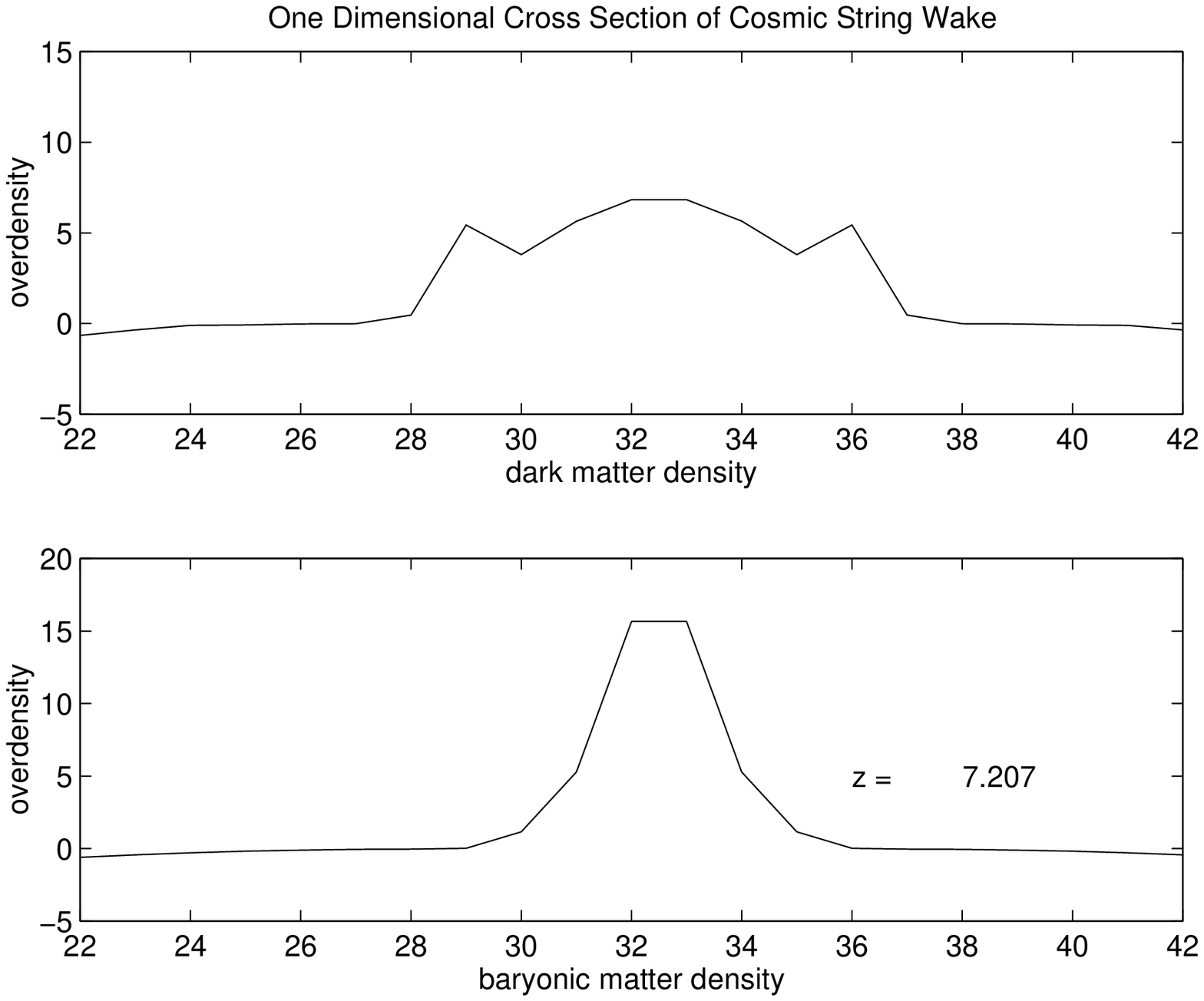]{}{}
\vfill\eject
Figures 1 - 6 depict the evolution in time of the matter overdensity
in dark matter and baryons of a planar symmetric cosmic string wake
formed at $z_i = 100$. The gridzone size is 0.0016 Mpc (remember the
figures are still at a relatively early stage in the evolution of the
wake, so the wake is still relatively thin in comoving coordinates).
\bigskip
\insertfig[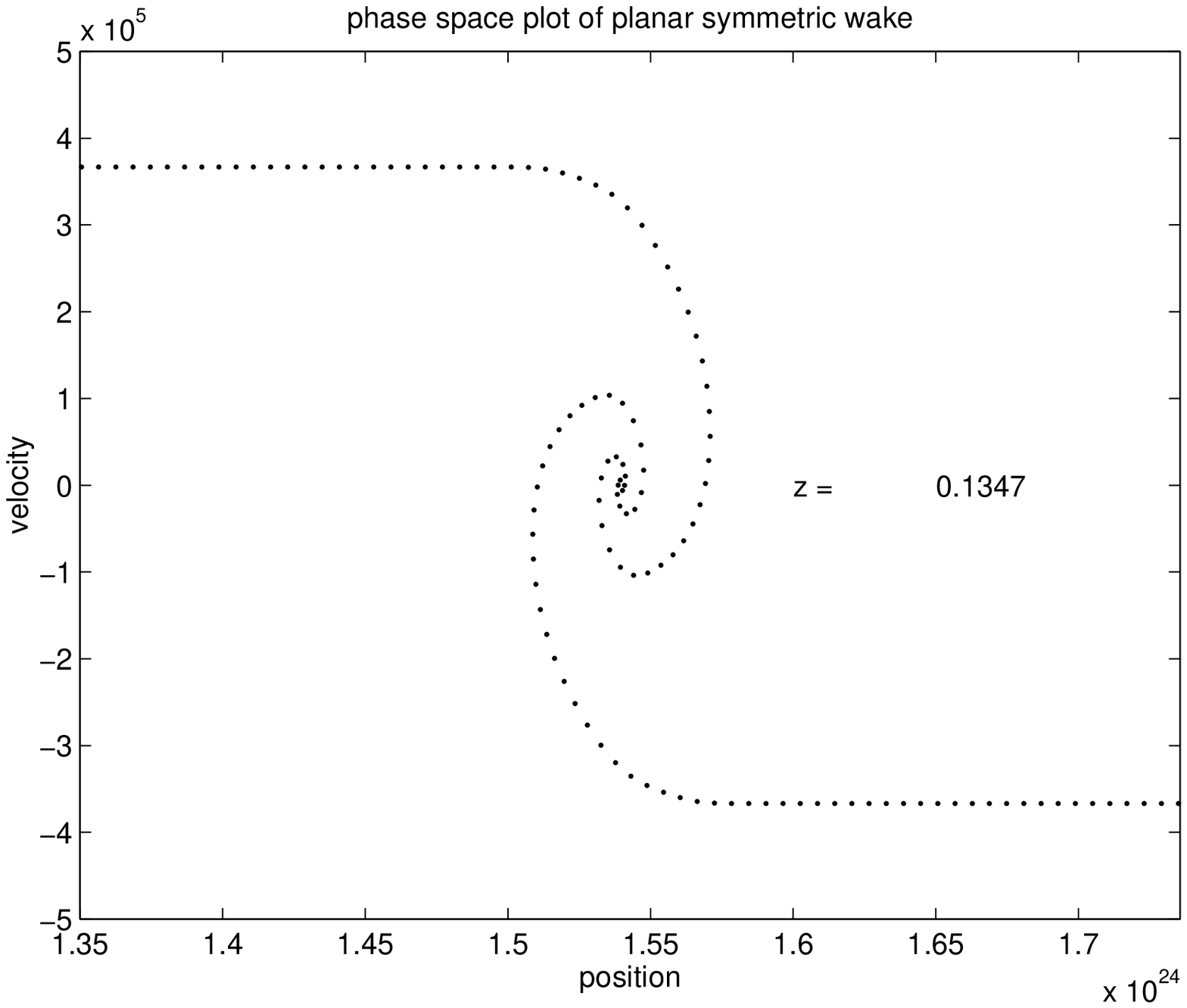]{This figure shows the phase space distribution of
dark matter particles accreting in a wake formed at $z_i =
100$. Position is in centimeters and velocity in centimeters/second}
\vfill\eject
\bigskip
\insertfig[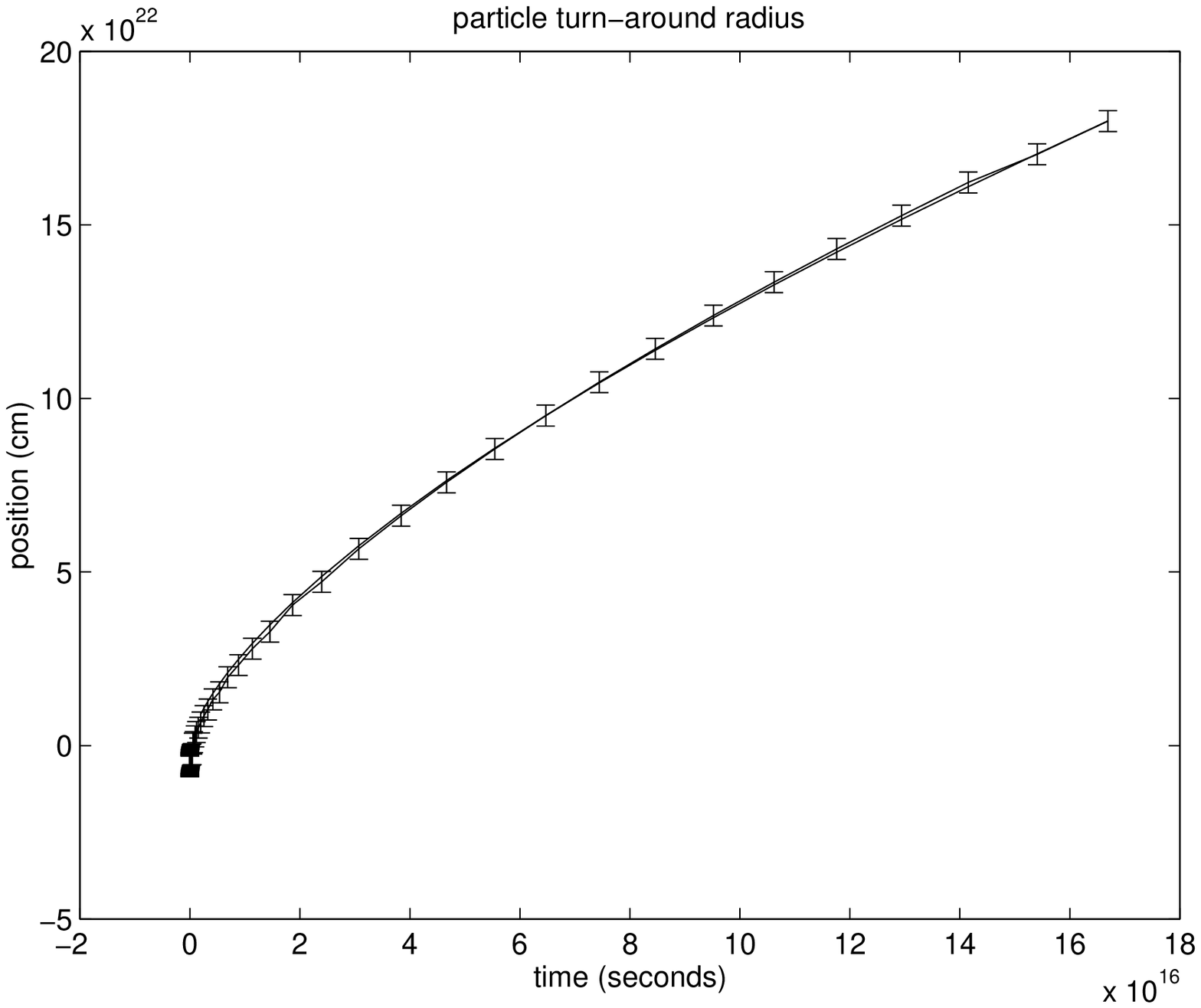]{In this figure the radius of secondary
turnaround (first caustic) is plotted vs. time. Also plotted is the
theoretical turnaround. Errorbars are the width of 1 gridzone. This
data is taken from a high-resolution simulation with 2048 gridzones
with $z_i = 10000$.}
\vfill\eject
\bigskip
\insertfig[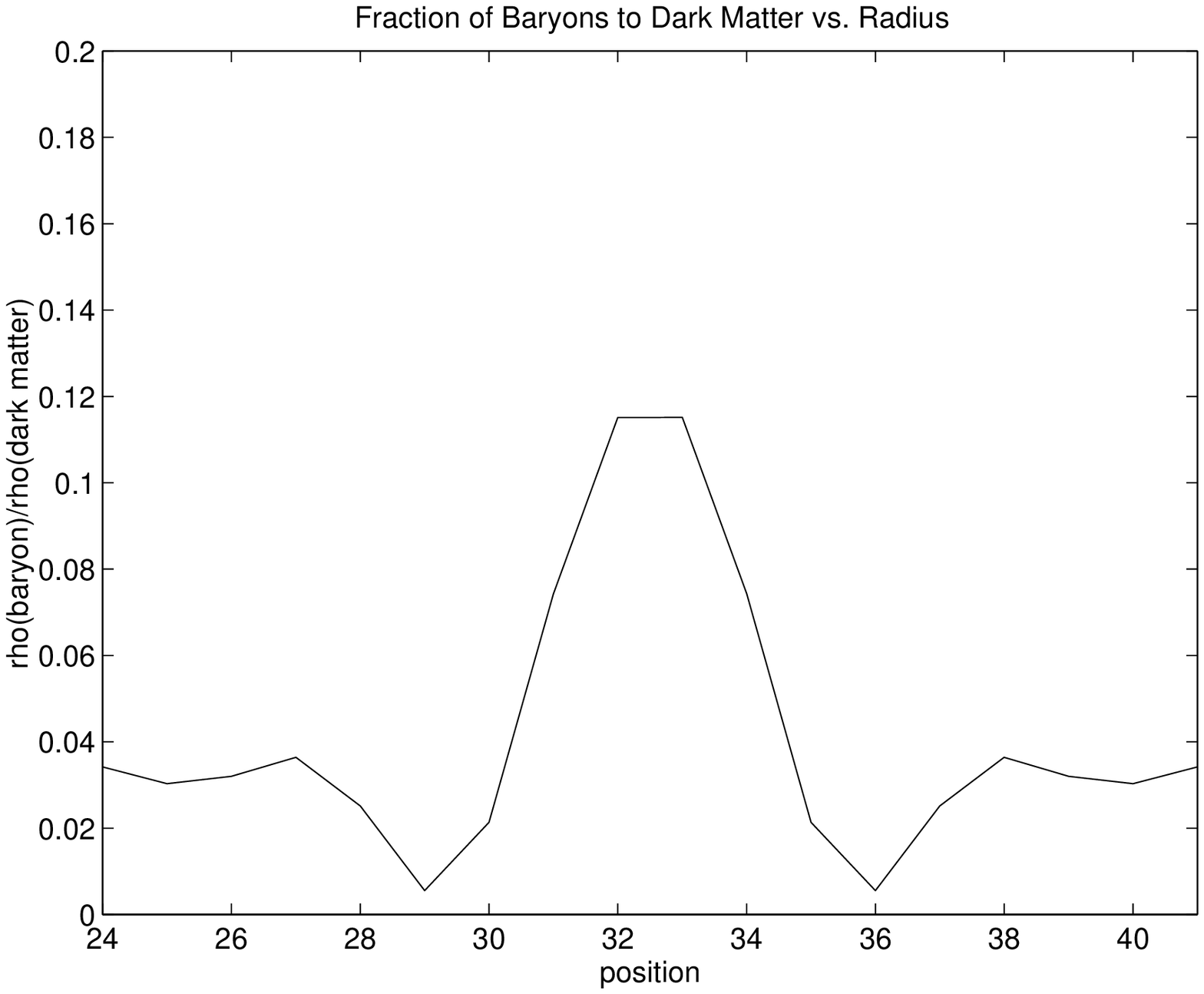]{The ratio of baryonic to dark matter at $z \sim
8$ in a planar symmetric cosmic string wake formed at $z_i = 100$. The
gridsize is 0.0016 Mpc.}
\vfill\eject
\inserttwofigs[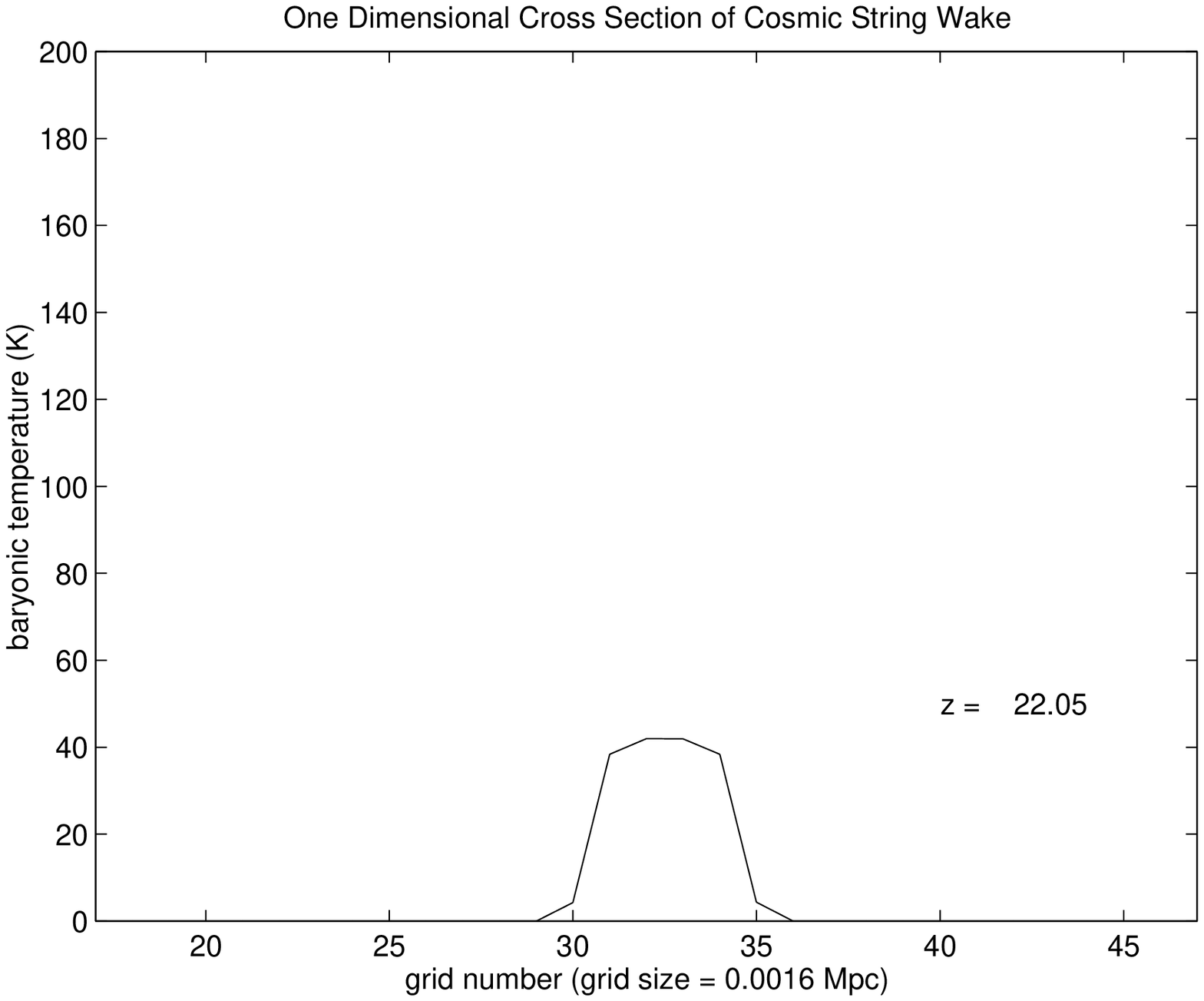,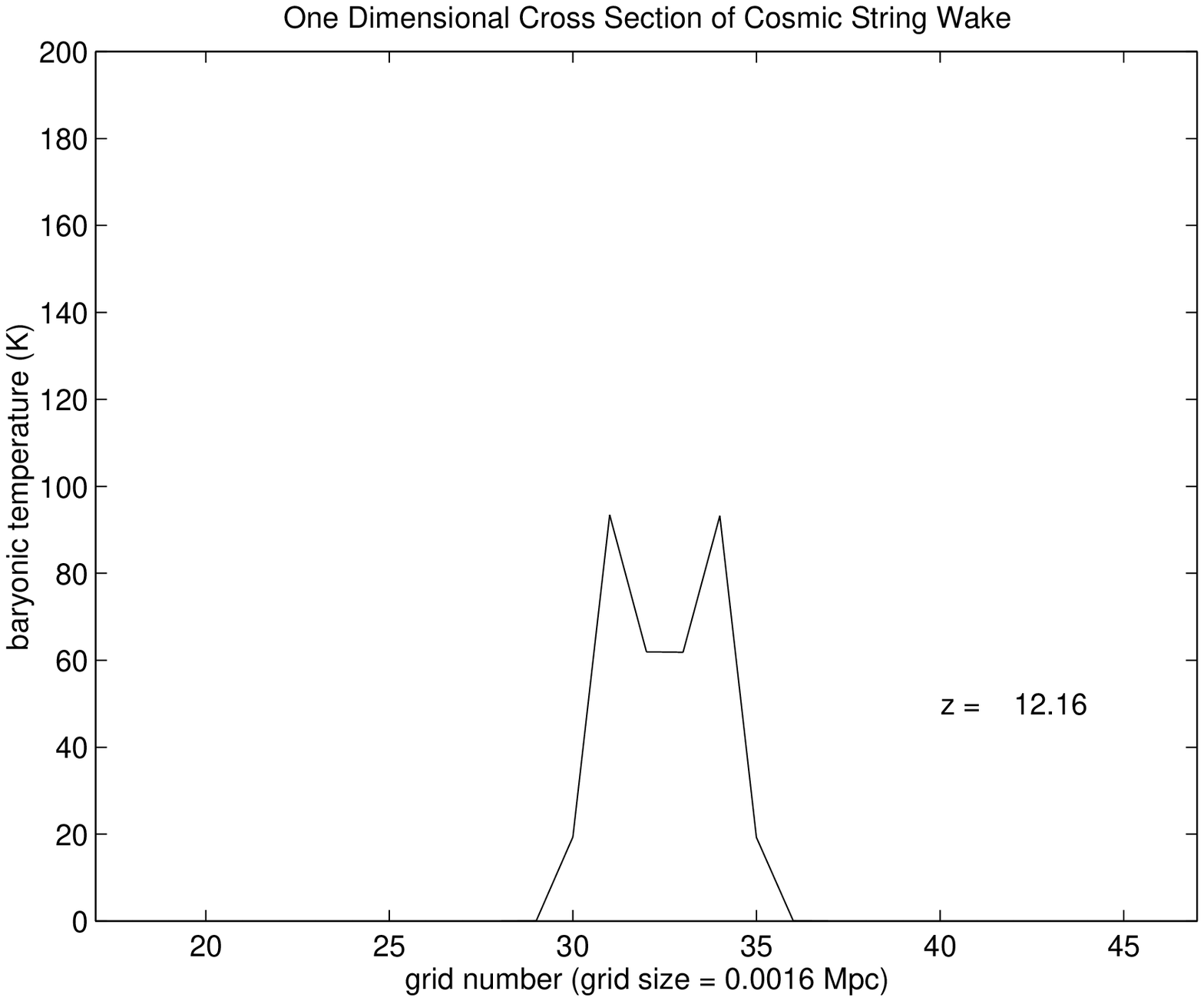]{}{}
\inserttwofigs[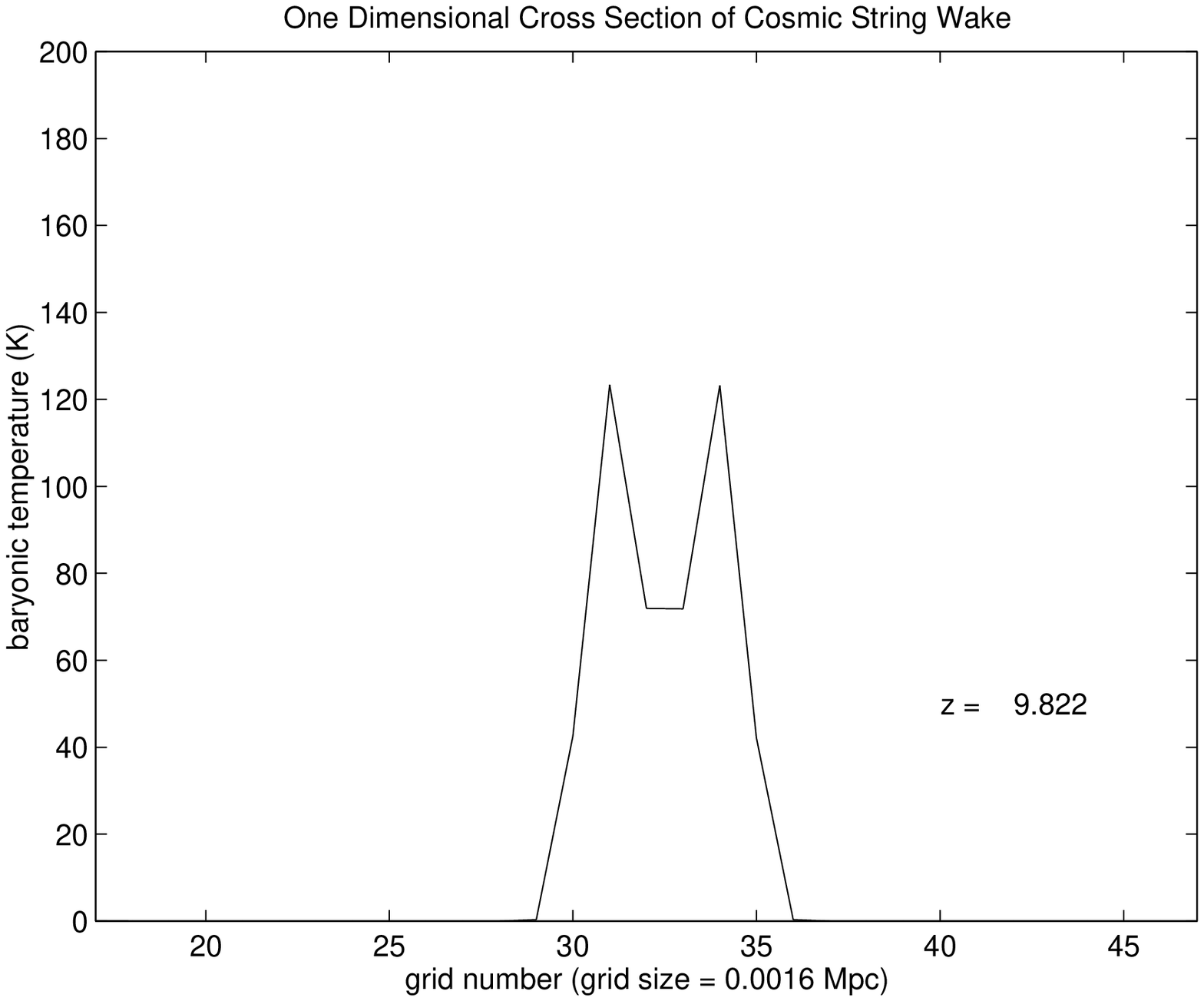,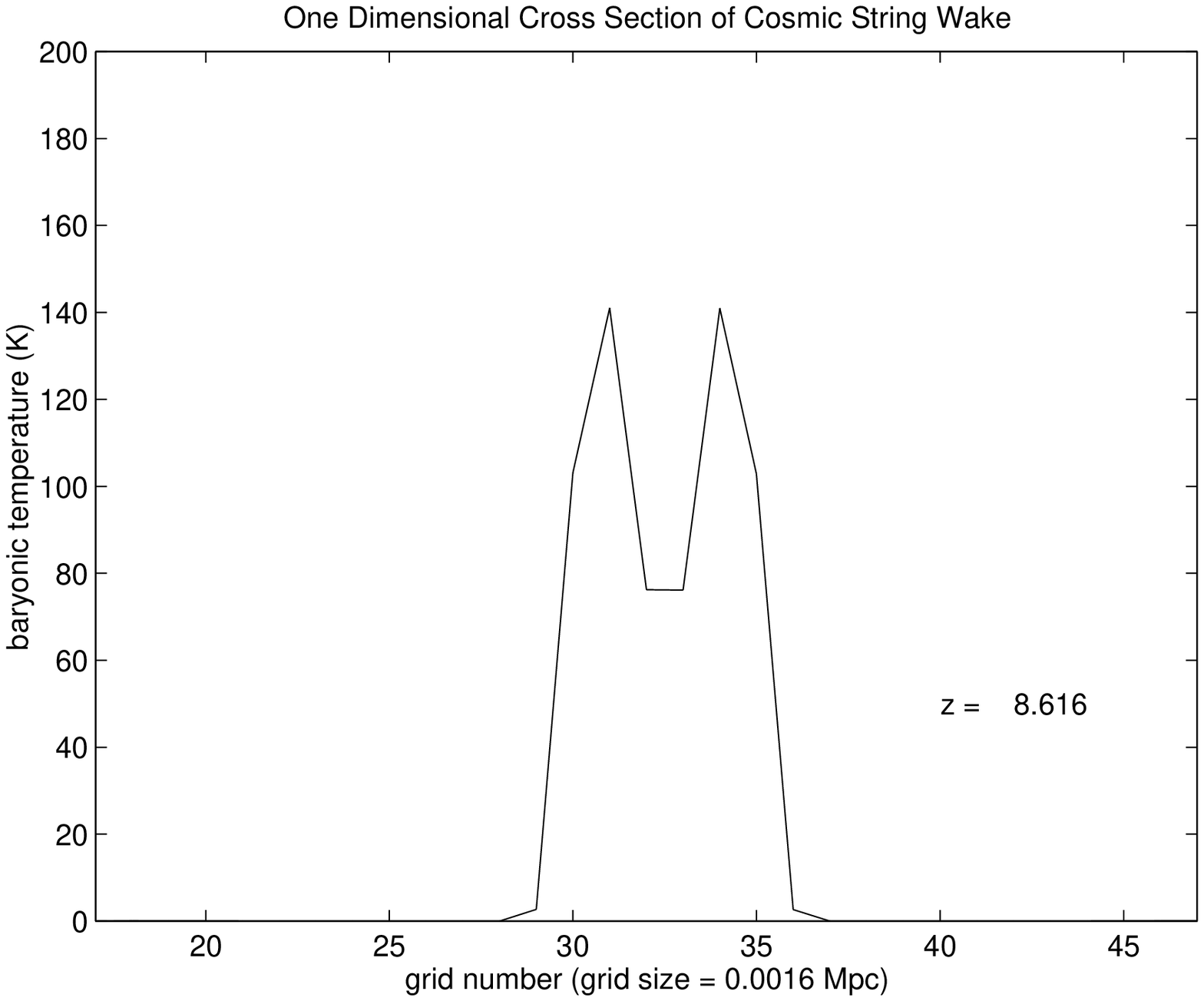]{}{}
\inserttwofigs[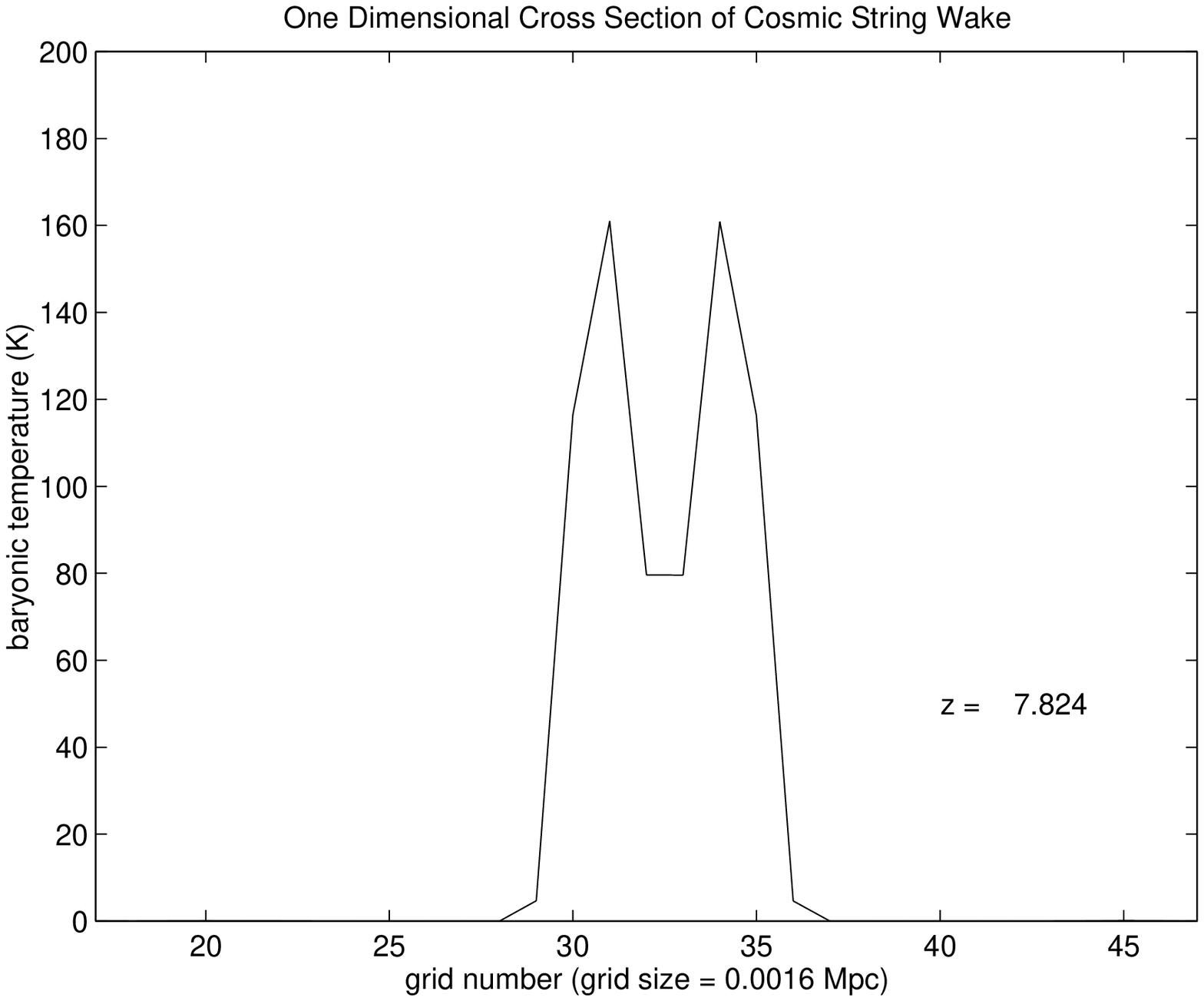,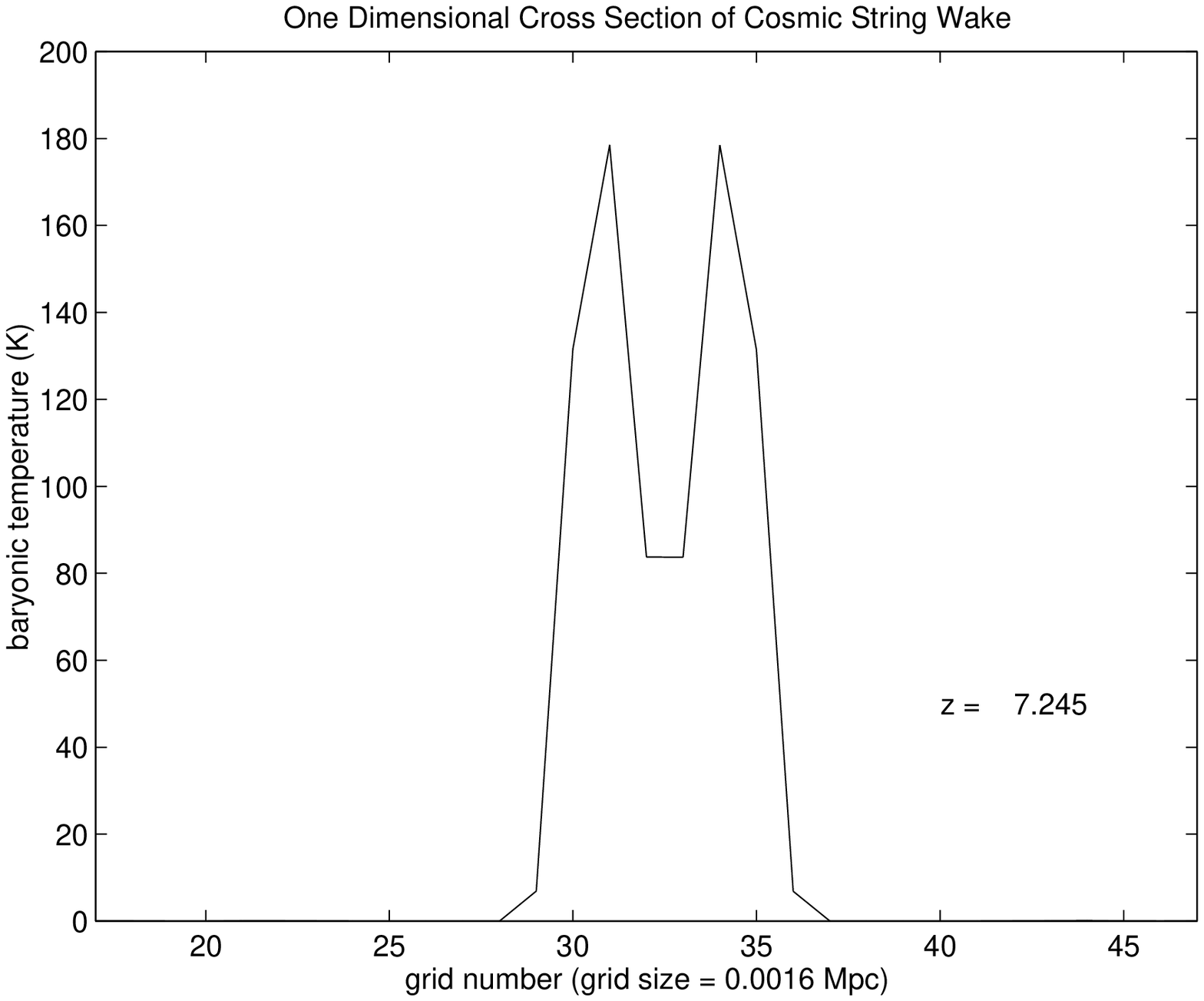]{}{}
\vfill\eject
Figures 10 - 15 depict the evolution in time of the temperature of the
baryonic matter in a planar symmetric cosmic string wake formed at
$z_i = 100$.
\bigskip
\insertfig[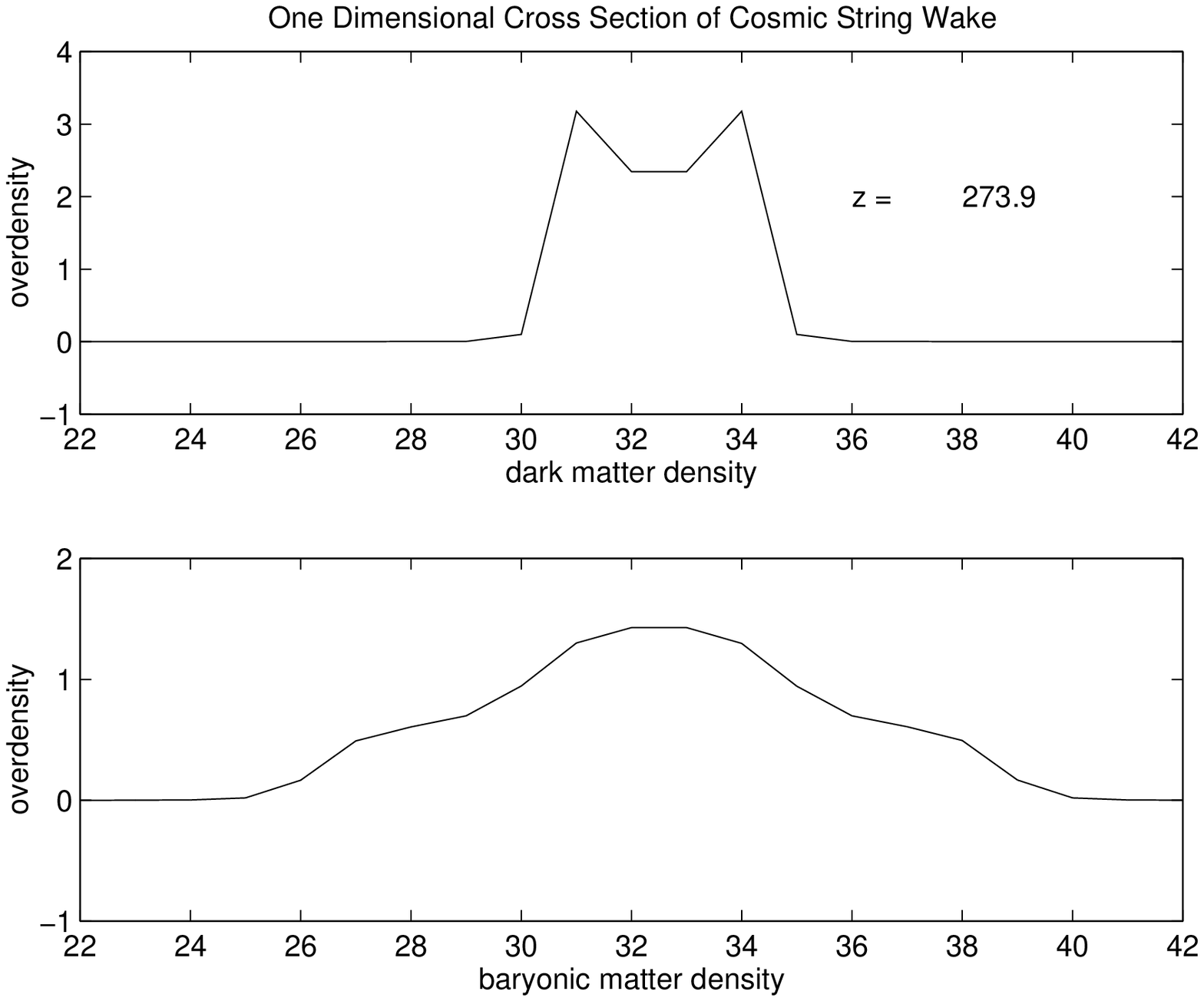]{This figure shows the dark and baryonic
matter distributions for a wake formed at $z_i = 1200$. For such early
formed wakes, the initial sound speed is high and a weak shock forms
spreading the baryonic overdensity beyond the width of the dark
matter. Later, as the gravity of the dark matter begins to dominate,
the baryonic matter clumps at the core of the wake.}
\vfill\eject
\bigskip
\insertfig[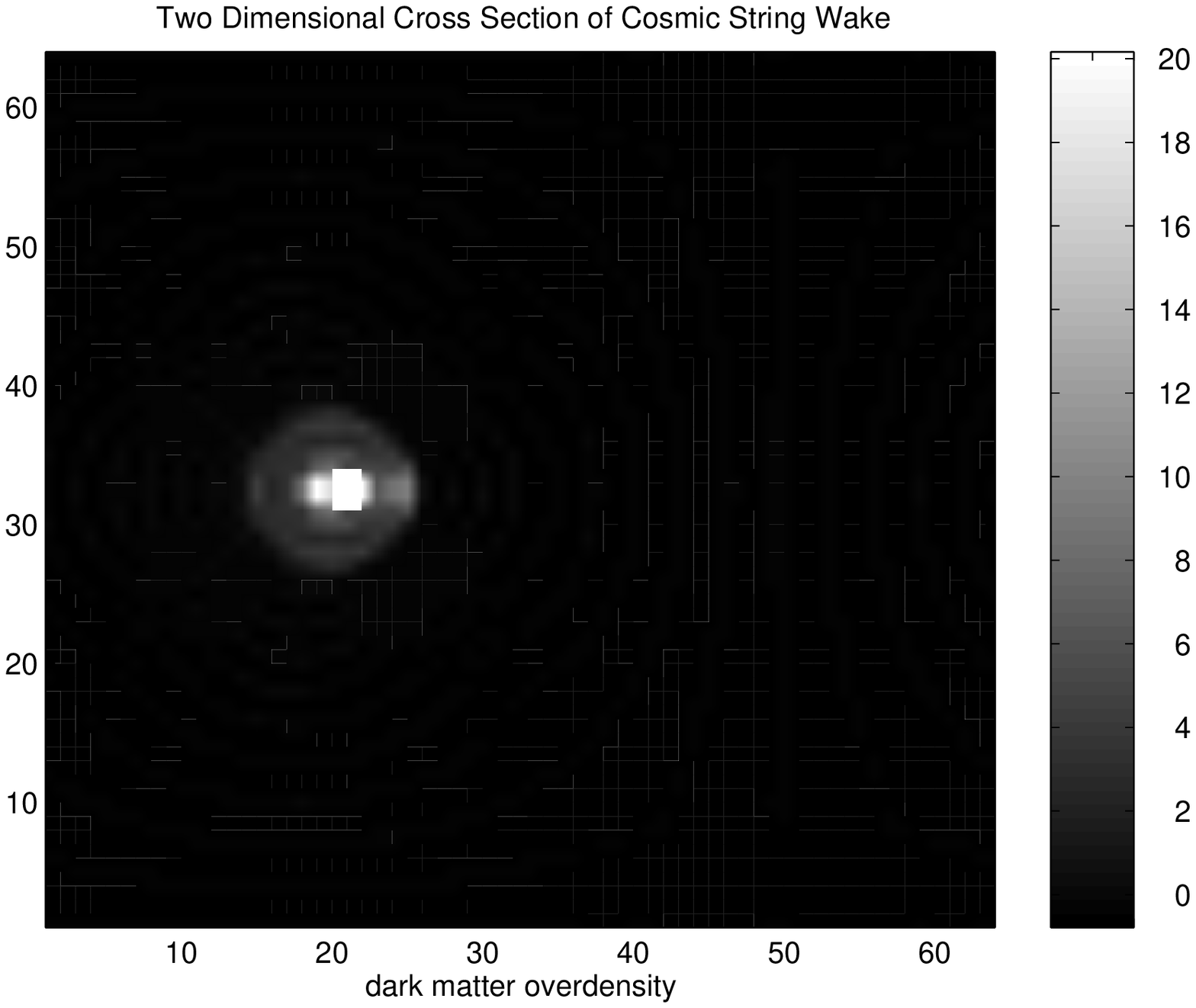]{The dark matter overdensity at $z \sim 8$
for a cosmic string filament formed at $z_i = 100$. The white square
is where pixels containing the high density from the string were
removed. The volume of the simulation is 1 Mpc x 1 Mpc.}
\vfill\eject
\bigskip
\insertfig[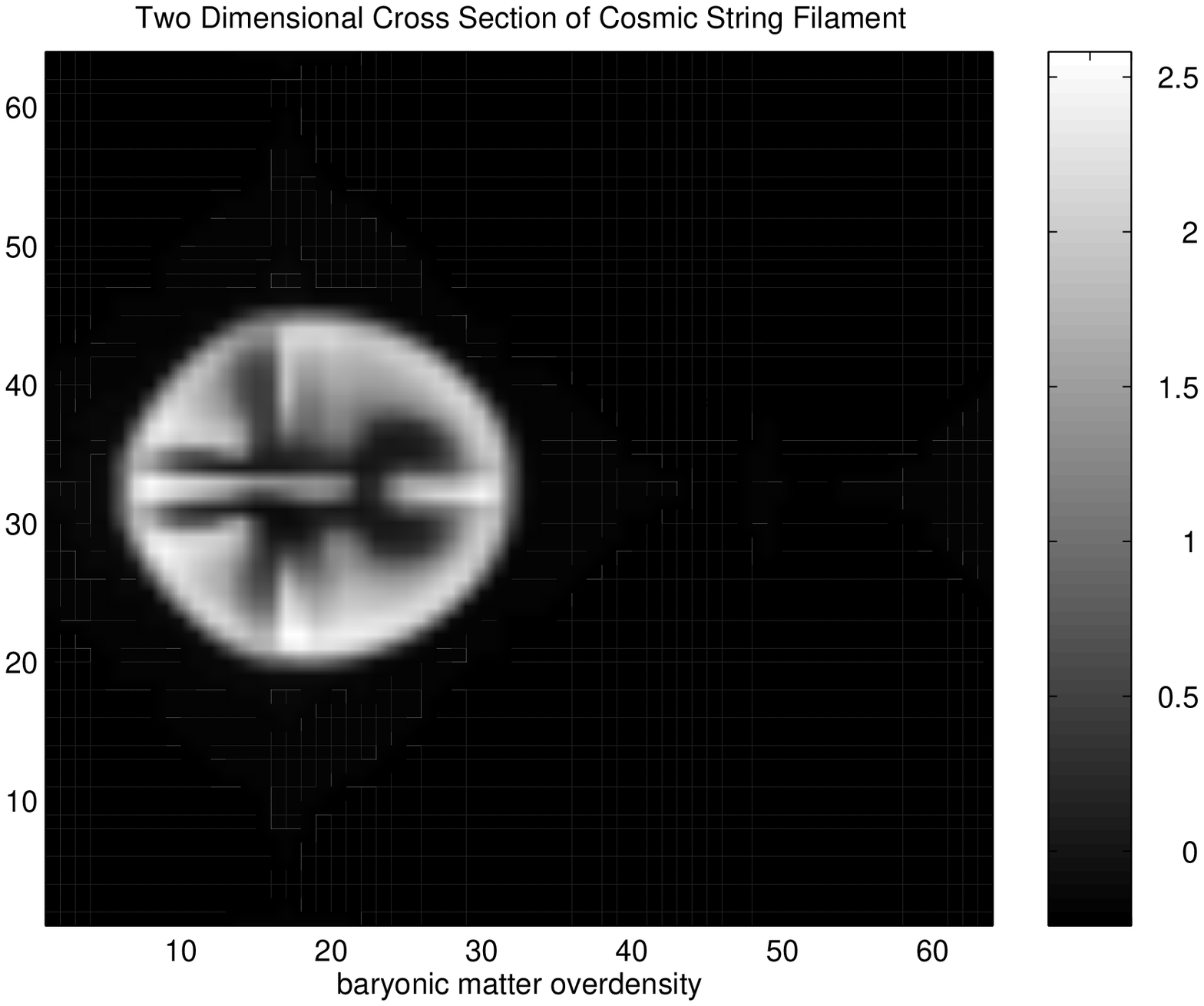]{The baryonic matter density at $z \sim 8$
for a cosmic string filament formed at $z_i = 100$. Notice the low
density in the core and the shell of matter that has been pushed out
of the core by high pressures. The volume of the simulation is 1 Mpc x
1 Mpc.}
\vfill\eject
\bigskip
\insertfig[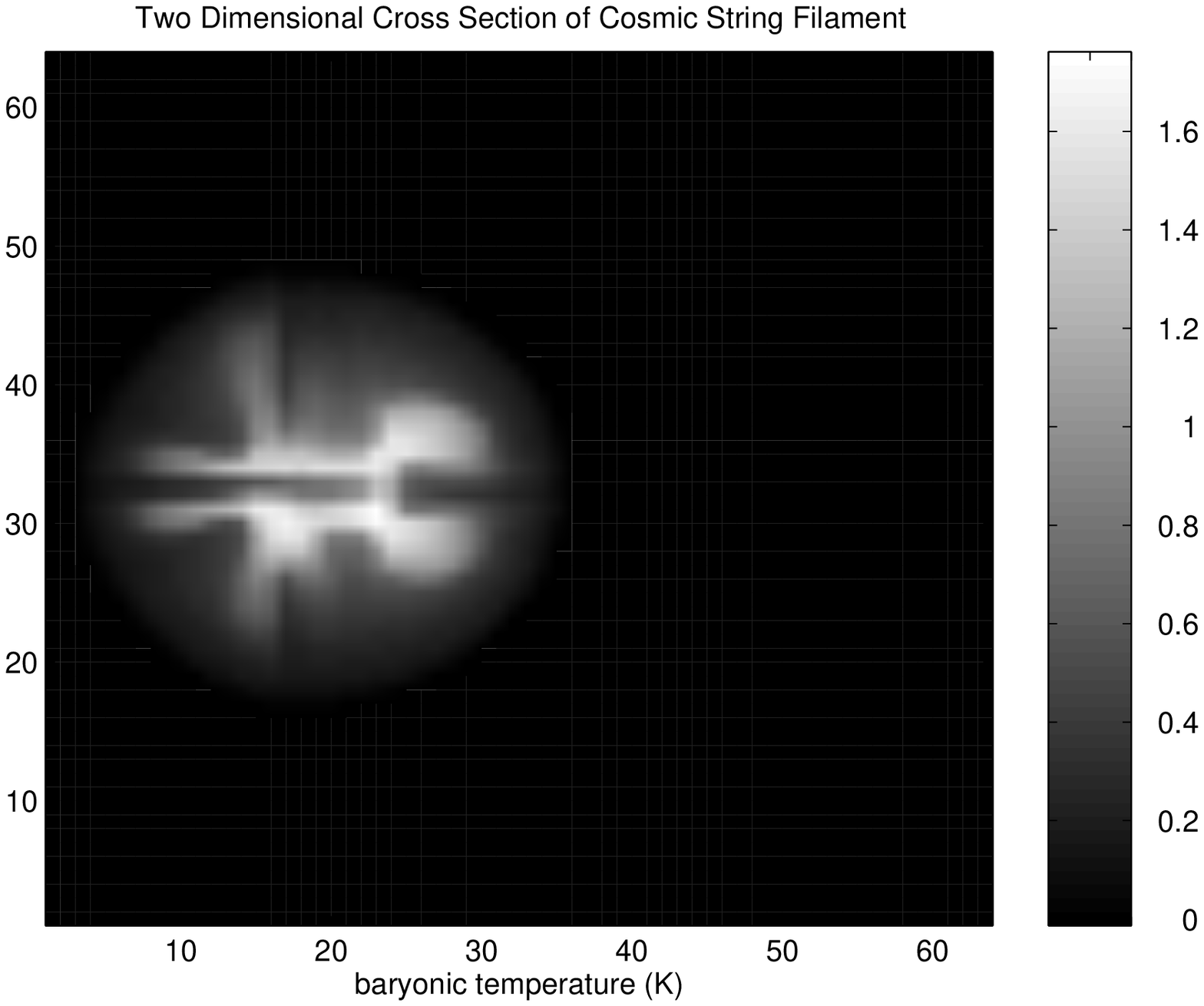]{The baryonic temperature at $z \sim 8$ for a
cosmic string filament formed at $z_i = 100$. Very high temperatures
are concentrated at the filament core. The volume of the simulation is
1 Mpc x 1 Mpc. The temperature is in units of $1 \times 10^6$K}
\vfill\eject
\end